\documentstyle[12pt,aaspp4]{article}
\def	\beq	{\begin{equation}}
\def	\eeq	{\end{equation}}
\font\boldsym=cmmib10
\def	\Angstrom	{\,{\rm \AA}}		% Angstrom
\def	\DG	{{\rm DG}}
\def	\G	{{\rm G}}
\def	\H	{{\rm H}}
\def	\HH	{{\rm H}_2}

\def	\K	{{\rm K}}
\def	\Mfac	{{M}}

\def	\ahat	{\hat{\bf a}}

\def	\bB	{{\bf B}}

\def 	\bE	{{\bf E}}

\def	\bGam	{{\bf \Gamma}}

\def	\bJ	{{\bf J}}

\def	\bQ	{{\bf Q}}

\def	\bk	{{\bf k}}
\def	\bmu	{{\hbox{\boldsym\char'026}}}	%bold \mu

\def	\bomega	{{\hbox{\boldsym\char'041}}}	%bold \omega

\def	\bphi	{{\hbox{\boldsym\char'036}}}	%bold \phi

\def	\bxi	{{\hbox{\boldsym\char'030}}}	%bold \xi
\def	\cm	{\,{\rm cm}}
\def	\eff	{{\rm eff}}
\def	\ehat	{\hat{\bf e}}
\def	\erg	{\,{\rm ergs}}
\def	\eV	{\,{\rm eV}\,}
\def	\g	{\,{\rm g}}
\def	\iso	{{\rm iso}}
\def	\khat	{\hat{\bf k}}
\def	\lambdabar	{\bar{\lambda}}
\def	\micron	{\mu{\rm m}}
\def	\nH	{n_{\rm H}}

\def	\phihat	{\hat{\bphi}}
\def	\s	{\,{\rm s}}
\def    \simlt  {\lower.5ex\hbox{$\; \buildrel < \over \sim \;$}}
\def    \simgt  {\lower.5ex\hbox{$\; \buildrel > \over \sim \;$}}
\def	\gtsim	{\simgt}
\def	\ltsim	{\simlt}
\def	\yr	{{\rm yr}}

\def	\xihat	{\hat{\bxi}}

\def	\science {{\it Science}}

\lefthead{Draine \& Weingartner}
\righthead{Radiative Torques on Interstellar Grains: Alignment}

\begin{document}
\author{{\bf POPe-689}.
\quad\quad\quad\quad\quad\quad\quad\quad\quad\quad\quad\quad\quad\quad\quad\quad
({\rm Submitted to} {\it Ap. J.})}
\title{Radiative Torques on Interstellar Grains:\\
	II. Grain Alignment}
\author{B.T. Draine}
\affil{Princeton University Observatory, Peyton Hall,
	Princeton, NJ 08544, USA; draine@astro.princeton.edu}

\and

\author{Joseph C. Weingartner}
\affil{Physics Dept., Jadwin Hall, Princeton University,
	Princeton, NJ 08544, USA; josephw@phoenix.princeton.edu}

\begin{abstract}
Radiative torques on irregular 
dust grains, in addition to producing superthermal rotation, 
play a direct
dynamical role in the alignment of interstellar dust with the local
magnetic field.
The equations governing the orientation of spinning, precessing grains
are derived;
$\HH$ formation torques and
paramagnetic dissipation are included in the dynamics.
Stationary solutions (constant alignment angle and spin rate) are found;
these solutions may be stable (``attractors'') or unstable (``repellors'').
The equations of motion
are numerically integrated for three exemplary
irregular grain geometries, exposed to anisotropic radiation with the
spectrum of interstellar starlight.
The resulting ``trajectory maps'' are classified as ``noncyclic'', 
``semicyclic'', or ``cyclic'', with examples of each given.

We find that radiative torques result in rapid grain alignment,
even in the absence of paramagnetic dissipation.
It appears that radiative torques due to starlight can account for
the observed alignment of interstellar grains with the Galactic
magnetic field.
\end{abstract}

\keywords{ISM: Dust, Extinction -- Polarization -- Scattering}

\section{Introduction}

The discovery of the polarization of starlight (Hall 1949; Hall \& Mikesell 1949;
Hiltner 1949a,b) revealed
not only that interstellar dust particles 
were nonspherical, but that some process
had brought about large-scale alignment of these grains.
The interstellar magnetic field was immediately suggested as 
a grain alignment
agent possessing the large-scale coherence required
by the polarization observations.

Initial investigations considered magnetic processes for bringing about
the grain alignment.
Both 
``compass-needle'' alignment of ferromagnetic grains (Spitzer \& Schatzman 1949;
Spitzer \& Tukey 1949, 1951) and paramagnetic dissipation
in spinning grains (Davis \& Greenstein 1949, 1950a,b, 1951) were
initially proposed.
When HI Zeeman splitting (Verschuur 1969) 
and dispersion and Faraday rotation toward pulsars (Woltjer 1970) indicated
interstellar field strengths of 3-4$\mu$G,
it was evident that
``compass-needle'' alignment was negligible,
but 
paramagnetic dissipation 
on the Davis-Greenstein timescale $\tau_\DG$ 
[see eq.(\ref{eq:t_DG}) below]
appeared to be viable, particularly if the
grains were superparamagnetic (Jones \& Spitzer 1967).
However, study of the statistical mechanics of grain alignment 
(Jones \& Spitzer 1967; Purcell \& Spitzer 1971) 
raised questions about the ability
of the interstellar magnetic field to achieve the observed degree of
grain alignment, since random gas-grain collisions (and magnetic fluctuations
within the grain) would tend to oppose the alignment process.

Martin (1971) pointed out that
because of the Rowland effect,
charged interstellar grains would have a substantial
magnetic moment either parallel or antiparallel to the angular velocity; 
this magnetic moment ensured that 
precession of the grain angular momentum around the magnetic field would be
sufficiently rapid that the observed interstellar grain alignment should always
be either parallel or perpendicular to the magnetic field direction,
regardless of the mechanism responsible for grain alignment.
Thus the galactic magnetic field could lead to large-scale coherence in
observed grain alignment, even if the magnetic field itself played no direct
role in grain alignment.
This conclusion was reinforced when Dolginov \& Mytrophanov (1976) noted
that the Barnett effect implied much larger magnetic moments for spinning
grains, leading to
even more rapid precession of $\bJ$ around $\bB_0$.

A very important development occurred when 
Purcell (1975, 1979) pointed out that grains were subject to systematic
torques which, in diffuse clouds, would result in grain rotation rates
far in excess of those which had been previously assumed to result from
simple thermal excitation by collisions with gas atoms.
Purcell considered three sources of systematic torques:
$\HH$ formation at preferred sites on the grain surface,
photoelectron emission following absorption of a UV photon,
and variations in the ``accommodation coefficient'' over the grain surface;
the $\HH$ formation torques appeared to be most important.
Purcell observed that if these torques were long-lived, then normal
paramagnetic relaxation would gradually bring the grain angular momentum
into alignment with the magnetic field.
However, if the grain surface changes on time scales short compared
to $\tau_\DG$, substantial grain alignment would not occur.
Spitzer \& McGlynn (1978) analyzed the ``crossover'' process when the
grain's rotation rate (in body coordinates) undergoes a reversal due
to a change in sign of the ``Purcell torque'', and obtained an estimate
for the disalignment caused by frequent crossovers.
Spitzer \& McGlynn concluded that if the active sites for $\HH$ formation
were short-lived, the disorientation during crossovers was too great for
grains to be aligned by ordinary paramagnetic dissipation.
The crossover phenomenon has recently been reanalyzed 
by Lazarian \& Draine (1997), who find that the disalignment per crossover
can be (paradoxically) substantially suppressed by 
``Barnett effect fluctuations'', 
with the result that paramagnetic
alignment could potentially explain the observed alignment of 
$a\gtsim0.1\micron$ grains, {\it if} no other systematic torques acted
to change the grain alignment.

In a previous paper (Draine \& Weingartner 1996a; hereinafter Paper I) 
we found that
interstellar starlight exerts
very substantial torques on irregular grains.
These ``radiative torques'' drive extreme superthermal rotation in
grains with the sizes ($a\approx0.2\micron$) which must be aligned to
account for the observed interstellar polarization.
In addition, radiative torques can act to change the direction of the
grain rotation, and must therefore play a role in the process of
grain alignment.
This paper is devoted to a study of this phenomenon.

The grain geometry and coordinate systems are introduced in
\S\ref{sec:coords}.
\S\ref{sec:torques} discusses the torques acting on the grain due to
the Barnett effect, gas drag, $\HH$ formation, paramagnetic dissipation,
and absorption and scattering of starlight.

In \S\ref{sec:dynamics} we obtain the equations of motion after
averaging over precession, and we discuss the stationary points and
``crossover points'' allowed by these equations.
``Trajectory maps'' showing the evolution of the grain orientation
and rotational velocity are discussed in \S\ref{sec:traj_maps}, where
we show that three different classes of grain behavior are possible.
Examples of each of the three classes of trajectory map are presented.
In many cases radiative torques, together with magnetic precession of $\bJ$
around $\bB_0$, rapidly bring about alignment of the
grain angular momentum $\bJ$ with the magnetic field $\bB_0$.

In \S\ref{sec:discuss} we summarize the behavior of our three
exemplary grains as a function of the angle $\psi$ between the magnetic
field $\bB_0$ and the starlight anisotropy direction.
Since the surface density of active sites for $\HH$ formation is
uncertain, we examine the sensitivity of our results to this quantity,
and find that, for likely values, radiative torques are more important than
paramagnetic dissipation for producing grain alignment.
In some cases the radiative torques do not allow a stationary solution;
even in this case the average alignment can be substantial.
Our results are summarized in \S\ref{sec:summary}.

%As a result, interest was renewed in other proposed alignment mechanisms,
%including alignment by gas-grain
%streaming (Gold 1952;
%Roberge \& Hanany 1990; Lazarian 1994, 1995a; 
%Roberge, Hanany \& Messinger 1995).
%However, the actual levels of gas-grain streaming do not appear to be
%sufficient for gas-grain streaming mechanisms to account for the
%observed widespread grain alignment.

\section{Grain Dynamics in a Magnetic Field\label{sec:coords}}
\subsection{Grain Geometry\label{subsec:geometry}}

As in Paper I, we consider irregular grains of density
$\rho$ and volume $V$, with an ``effective radius''
$a_\eff\equiv(3V/4\pi)^{1/3}=10^{-5}a_{-5}\cm$.
In the present paper we report results for three specific
grain shapes, shown in Fig. \ref{fig:shapes}.
Shape 1, which can be described as an assembly of 13 cubes, 
is the target geometry considered in Paper I.
Shape 2 can be described as an assembly of 11 cubes, with coordinates given
in Table 1.
Shape 3 can be described as the union of 5 overlapping spheres, with coordinates
and radii given in Table 2.

The moment of inertia tensor has eigenvalues $I_1\geq I_2\geq I_3$, with
principal axes $\ahat_1$, $\ahat_2$, and $\ahat_3$.\footnote{
	Axes $\ahat_1$ and $\ahat_2$ are obviously determined only to within
	a choice of sign.  We arbitrarily choose one of the two solutions for
	$\ahat_1$ and likewise for $\ahat_2$, and ``freeze'' these in body
	coordinates.  The axis $\ahat_3\equiv\ahat_1\times\ahat_2$.
	}
We define dimensionless parameters $\alpha_j$ by
\beq
I_j \equiv \alpha_j {2\over 5}\rho V a_\eff^2 ~~~.
\eeq
A sphere has $\alpha_j=1$; the irregular grain of Paper I has $\alpha_1=1.745$.
In Table \ref{tab:alphaj} we give $\alpha_j$ for shapes 2 and 3.

The thermal rotation rate for the grain is
\beq
\omega_T\equiv \left({15\over 8\pi\alpha_1}{kT\over\rho a^5}\right)^{1/2}
= 1.66\times10^5 \left(T\over 100\K\right)^{1/2}
\left({3\g\cm^{-3}\over\rho}\right)^{1/2} \alpha_1^{-1/2}
a_{-5}^{-2.5} \s^{-1}
\eeq
for rotation around $\ahat_1$ with kinetic energy $kT/2$.

\subsection{Coordinate System}

\begin{figure}
\epsscale{0.50}
\plotone{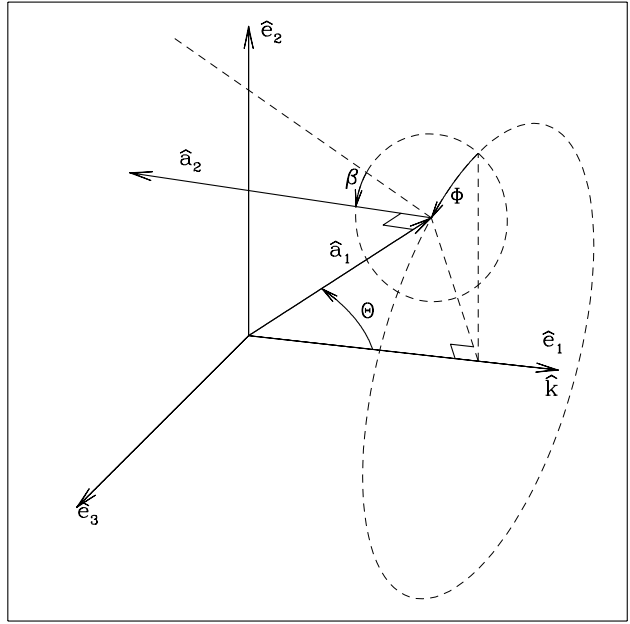}
\caption{
	\label{fig:ThetaPhi}
	Grain orientation showing ``scattering coordinates'',
	$\Theta$, $\Phi$, and $\beta$ characterizing the grain
	orientation relative to the incident radiation.
	$\khat$ is the direction of propagation of the incident
	radiation.
	}
\end{figure}

\begin{figure}
\epsscale{0.50}
\plotone{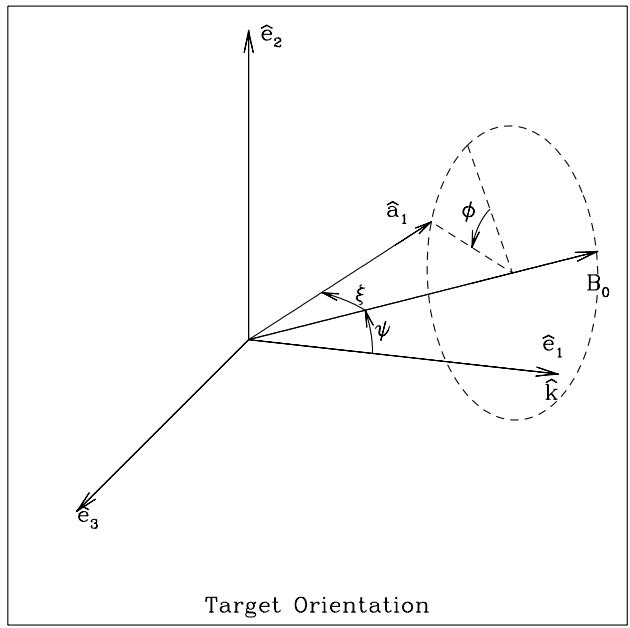}
\caption{
	\label{fig:xiphi}
	Grain orientation showing ``alignment coordinates''
	$\xi$ and $\phi$.
	$\bB_0$ is the static magnetic field, and $\psi$ is the
	angle between $\bB_0$ and $\khat$.
	}
\end{figure}

Consider unidirectional radiation, propagating in direction $\khat$.
Without loss of generality we may choose the unit vectors 
$\ehat_1$, $\ehat_2$, $\ehat_3$ defining
the ``scattering coordinates'' such that 
$\ehat_1$ is parallel to $\khat$, and the magnetic field $\bB_0$ lies
in the $\ehat_1,\ehat_2$ plane (see Figures \ref{fig:ThetaPhi},
\ref{fig:xiphi}).
Let $\psi$ be the angle between $\bB_0$ and $\khat$; since the sign of
$B_0$ does not affect the grain dynamics, we need only consider
$\psi\in[0,\pi/2]$.
Let the ``alignment angle'' 
$\xi\in[0,\pi]$ be the angle between the grain principal
axis $\ahat_1$ and $\bB_0$.
The superthermally-rotating grain will be spinning around $\ahat_1$;
because of the Barnett moment (see \S\ref{sec:barnett}), 
the axis $\ahat_1$ of the spinning grain
will rapidly precess in a cone around $\bB_0$, and it is convenient
to define a ``precession frame'' using polar coordinates $(r,\xi,\phi)$ 
with $\bB_0$ as the polar axis.
Then $(\xi,\phi)$ corresponds to the direction of the
grain axis $\ahat_1$:
\beq
\ahat_1 =
\ehat_1(\cos\psi\cos\xi-\sin\psi\sin\xi\cos\phi) +
\ehat_2(\sin\psi\cos\xi+\cos\psi\sin\xi\cos\phi) +
\ehat_3\sin\xi\sin\phi ~~,~~
\eeq
and unit vectors in the directions of increasing $\xi$ and $\phi$ are
given by
\begin{eqnarray}
\xihat \!&\!=\!&\!
-\ehat_1(\sin\psi\cos\xi\cos\phi+\cos\psi\sin\xi)
+\ehat_2(\cos\psi\cos\xi\cos\phi-\sin\psi\sin\xi)
+\ehat_3\cos\xi\sin\phi~,~~
\\
\phihat \!&\!=\!&\! 
\ehat_1\sin\psi\sin\phi - \ehat_2\cos\psi\sin\phi + \ehat_3\cos\phi ~~~~.
\end{eqnarray}
The orientation angles $\Theta$ and $\Phi$ in the scattering frame 
(see Fig.\ \ref{fig:ThetaPhi}) can be determined from
\beq
\Theta(\xi,\phi)=\cos^{-1}\left[\cos\psi\cos\xi-\sin\psi\sin\xi\cos\phi\right]
~~~~,
\label{eq:Theta}
\eeq
\beq
\Phi(\xi,\phi)=2\tan^{-1}\left[
{\sin\Theta-\sin\psi\cos\xi-\cos\psi\sin\xi\cos\phi
\over \sin\xi\sin\phi}
\right]
~~~~.
\label{eq:Phi}
\eeq
The grain orientation is also determined by the angle $\beta$ measuring
rotation of the grain around $\ahat_1$ (see 
Paper I and Fig.\ \ref{fig:ThetaPhi}).
In the present study we will assume the grain to spin rapidly around
$\ahat_1$ (see below) and will therefore employ radiative torques
obtained after averaging over angle $\beta$; henceforth, this angle will
not be explicitly mentioned.

\subsection{Grain Dynamics}
When $\omega^2\gg\omega_T^2$, internal dissipation in the grain will
tend to bring the
principal axis of largest
moment of inertia $\ahat_1$ into alignment with the angular moment $\bJ$,
as this is the state of minimum kinetic energy for fixed $\bJ$.
As discussed by Purcell (1979), both viscoelasticity and Barnett
relaxation are effective.
In fact, thermal fluctuations will act to prevent perfect alignment of
$\bJ$ with $\ahat_1$
(Lazarian \& Roberge 1997).
In the analysis below we will assume perfect alignment of $\ahat_1$ with
$\bJ$, so that $\bomega=\omega\ahat_1$.
Then
\beq
I_1\left[
\ahat_1{d\omega\over dt}  + 
\xihat\omega{d\xi\over dt}+
\phihat\omega\sin\xi{d\phi\over dt}
\right]
= \bGam_{\rm B} + \bGam_{rad} +  \bGam_{\HH} + \bGam_{drag} + \bGam_\DG
~~~~,
\label{eq:eqofmotion}
\eeq
where $\bGam_{\rm B}$ is the torque due to the Barnett moment,
$\bGam_{rad}$ is the radiative torque due to starlight,
$\bGam_{\HH}$ is the torque due to $\HH$ formation,
$\bGam_{drag}$ is the drag torque due to gas atoms (and photon emission),
and
$\bGam_\DG$ is the ``Davis-Greenstein'' torque due to paramagnetic
dissipation.
These torques are discussed below.

\section{Torques\label{sec:torques}}

\subsection{Barnett Moment and Precession around $\bB_0$\label{sec:barnett}}

Spinning grains develop a magnetic moment $\bmu$ antiparallel to $\bomega$,
primarily due to the Barnett effect (Dolginov \& Mytrophanov 1976); if
the grain is charged there is a small correction due to the Rowland
effect which we shall neglect.
The Barnett magnetic moment is 
\beq
\bmu = -{\chi(0)V\hbar \over g \mu_B}\bomega ~~~~,
\eeq
where $V$ is the grain volume, $\mu_B$ is the Bohr magneton,
$g\approx2$ is the gyromagnetic ratio, 
and $\chi(0)$ is the static susceptibility.
The magnetic properties of interstellar grains have recently been
reviewed by Draine (1996); ``normal'' paramagnetism
is expected to result in $\chi(0)\approx10^{-4}$, and
superparamagnetism would give even larger values of
$\chi(0)$.

The torque resulting from the 
Barnett effect magnetic moment is
\beq
\bGam_{\rm B} = \bmu\times\bB_0 = 
-\phihat I_1 \Omega_{\rm B}\omega\sin\xi ~~~.
\eeq
This torque will cause the grain to precess in the Galactic magnetic field
$\bB_0$ with a precession frequency
\beq
\label{eq:omegab}
\Omega_B = {\mu B_0 \over I_1 \omega}
= {5\hbar\chi(0)B_0 \over 2\alpha_1 g \mu_B \rho a_\eff^2}
\approx
7.5\yr^{-1} a_{-5}^{-2}\left({3\g\cm^{-3}\over\alpha_1\rho}\right)^{1/2}
\left({\chi(0)\over10^{-4}}\right)
\left({B_0\over 5\mu\G}\right) ~~~~,
\eeq
where we have set $g\approx2$.
It is therefore clear that interstellar grains will precess around $\bB_0$
very rapidly compared to all other timescales except the grain rotation
period itself.

\subsection{Gas Drag}

The drag torque is
\beq
\bGam_{drag} = -\ahat_1~I_1~\omega~\tau_{drag}^{-1} ~~~,
\eeq
where the timescale for gas drag is (Paper I)
\beq
\tau_{drag} = {\pi~\alpha_1~\rho~ a_\eff \over
3\delta\nH (2\pi m_\H kT)^{1/2}}
= 8.74\times10^4\yr {\alpha_1\over\delta}
\left(\rho\over3\g\cm^{-3}\right)~ a_{-5} T_2^{1/2}
\left({3000\cm^{-3}K\over\nH T}\right) ~~~,
\label{eq:t_drag}
\eeq
where the drag coefficient $\delta$ is of order unity.
A sphere has $\delta=1$; we estimate $\delta\approx2$ for shapes 1 and 2,
and $\delta\approx1.5$ for the more compact shape 3.

There is also drag on the grain due to emission of far-infrared radiation;
while it can be comparable to gas drag for very small grains (Paper I),
it is unimportant for the $a_\eff\approx0.1\micron$ grains
we shall be interested in here, so we shall neglect it.

\subsection{H$_2$ Formation\label{sec:H2form}}

Purcell (1979) estimated the torque due to $\HH$ formation on the grain
surface, after averaging over the grain rotation, to be
\beq
\label{eq:gamma_h2}
\bGam_{\HH}\cdot\ahat_1 = {1\over3}
\left({\pi\over3}\right)^{1/6} f~ n(\H) ~(2E kT)^{1/2} a_\eff^2 ~l~ p(t) ~~~.
\eeq
where $l^2$ is the surface area per $\HH$ formation site on the
grain surface,
$f$ is the fraction of the arriving H atoms which depart as $\HH$,
$E$ is the kinetic energy of the departing $\HH$ molecules,
and $p(t)$ is a random variable with time averages
$\langle p(t)\rangle=0$,
$\langle p(t)p(t+\tau)\rangle=e^{-\tau/t_L}$,
where $t_L$ is the ``lifetime'' of a surface recombination site.
We define a characteristic rotation rate
\begin{eqnarray}
\omega_{\HH}
&=&{5\over24}
\left({3\over\pi}\right)^{1/3}
{f\over\delta}
\left({E\over3m_\H}\right)^{1/2}
\left({n(\H)\over\nH}\right)
{l\over a_\eff^2}
\\
&\approx& 5.20\times10^7\s^{-1}
{f\over \delta a_{-5}^2}
\left({l\over10\Angstrom}\right)
\left({E\over 0.2\eV}\right)^{1/2}
\left({n(\H)\over\nH}\right)
\label{eq:omegah2}
\end{eqnarray}
such that the $\HH$ torque, averaged over the grain rotation, is
\beq
\bGam_{\HH} = \ahat_1{I_1 \omega_{\HH}\over \tau_{drag}} p(t) ~~~.
\label{eq:gamma_h2b}
\eeq
Note that $\omega_{\HH}$ is proportional to $l$, the characteristic
separation between $\HH$ formation sites.
We will usually assume $l=10\Angstrom$ for purposes of illustration,
but other values of $l$ will be considered below.

\subsection{Paramagnetic Dissipation}

Magnetic dissipation in the grain produces a torque
(Davis \& Greenstein 1951; Jones \& Spitzer 1967)
\beq
\bGam_\DG = -(\xihat\sin\xi\cos\xi + \ahat_1\sin^2\xi)
I_1\omega~\tau_\DG^{-1}
\eeq
where
\beq
\tau_\DG 
= 
{2\alpha_1\rho a_\eff^2 \over5K(\omega) B_0^2}
=
1.5\times10^6\yr~\left({\alpha_1\rho\over3\g\cm^{3}}\right) a_{-5}^2 
\left( {10^{-13}\s\over K(\omega)}\right)
\left( {5\mu\G\over B_0}\right)^2 ~~~;
\label{eq:t_DG}
\eeq
$K(\omega)\equiv\chi^{\prime\prime}(\omega)/\omega$,
where $\chi^{\prime\prime}$ is the imaginary part of the
complex susceptibility.
We expect
$K\approx 10^{-13}\s$ for normal paramagnetism and
$\omega\ltsim10^9\s^{-1}$ (Jones \& Spitzer 1967; Draine 1996).

\subsection{Radiative Torques}

As in Paper I, we approximate the interstellar radiation field in a
diffuse cloud by 
an isotropic component with energy density $(1-\gamma)u_{rad}$ 
plus a unidirectional component
with energy density $\gamma u_{rad}$.
The radiative torque can be written (cf. Paper I)
\beq
\bGam_{rad}(\xi,\phi) =
{u_{rad} a_\eff^2\lambdabar \over 2}
\left(
\gamma
\left[ 
F(\xi,\phi)\xihat + G(\xi,\phi)\phihat + H(\xi,\phi)\ahat_1
\right]
+ (1-\gamma)\langle Q_\Gamma^\iso\rangle\ahat_1
\right)   ~~~,
\eeq
\begin{eqnarray}
F(\xi,\phi)
&=&
\langle\bQ_\Gamma\rangle\!\cdot\!\ehat_1(-\sin\psi\cos\xi\cos\phi-\cos\psi\sin\xi) +
\nonumber
\\
&&
\langle\bQ_\Gamma\rangle\!\cdot\!\ehat_2(\cos\psi\cos\xi\cos\phi-\sin\xi\sin\psi)
+\langle\bQ_\Gamma\rangle\!\cdot\!\ehat_3\cos\xi\sin\phi
\label{eq:F(xi,phi)}
\\
G(\xi,\phi)
&=&
\langle\bQ_\Gamma\rangle\!\cdot\!\ehat_1\sin\psi\sin\phi -
\langle\bQ_\Gamma\rangle\!\cdot\!\ehat_2\cos\psi\sin\phi + 
\langle\bQ_\Gamma\rangle\!\cdot\!\ehat_3\cos\phi
\label{eq:G(xi,phi)}
\\
H(\xi,\phi) &=&
\langle\bQ_\Gamma\rangle\!\cdot\!\ehat_1(\cos\psi\cos\xi-\sin\psi\sin\xi\cos\phi) +
\nonumber
\\
&&
\langle\bQ_\Gamma\rangle\!\cdot\!\ehat_2(\sin\psi\cos\xi+\cos\psi\sin\xi\cos\phi) +
\langle\bQ_\Gamma\rangle\!\cdot\!\ehat_3\sin\xi\sin\phi
\label{eq:H(xi,phi)}
\end{eqnarray}
where the radiative torque efficiency vector $\bQ_\bGam(\Theta,\Phi)$ 
depends on $\xi$, $\phi$, and
the angle $\psi$ between $\bB_0$ and $\bk$, 
$Q_\Gamma^\iso$ is the torque efficiency factor for
isotropic radiation,
and $\langle\rangle$ denotes averaging over the incident radiation spectrum,
here assumed to be the interstellar radiation field (``ISRF'') 
spectrum 
(Mezger, Mathis, \& Panagia 1982; Mathis, Mezger, \& Panagia 1983),
for which $u_{rad}=8.64\times10^{-13}\erg\cm^{-3}$,
$\bar{\lambda}=1.202\micron$.
$\langle Q_\Gamma^\iso\rangle_{ISRF}$ is given in Table \ref{tab:qfactors} for
shapes 1--3; we see that it is only $\sim1\%$ of 
$\ahat_1\cdot\langle\bQ_{\Gamma}\rangle_{ISRF} (\Theta=0)$, showing that
the radiative torque due to isotropic starlight is negligible compared to that
due to anisotropic starlight, unless the anisotropic component is less than
$\sim1\%$ of the total background.
The angles $\Theta(\xi,\phi)$ and $\Phi(\xi,\phi)$ are obtained using
eq.~(\ref{eq:Theta},\ref{eq:Phi}).
The grains are assumed to be composed of material
with the dielectric function of ``astronomical silicate'' (Draine \& Lee 1984).

As discussed in Paper I, it is sufficient to compute
only $\bQ_\Gamma(\Theta,0)$.
Expressions for $F$,
$G$, and $H$ in terms of $\bQ_\Gamma(\Theta,0)$ are given in the Appendix.
$\bQ_\Gamma(\Theta,0)$ 
was computed as described in Paper I, assuming $\Phi=0$ and a number of
different values of the grain rotation angle $\beta$.
%$Q(\Theta,\Phi)$ was computed from $Q(\Theta,0)$ as described in Paper I.

As discussed in Paper I, computation of $\bQ_\Gamma(\Theta,0)$ as a function
of both $\Theta$ and wavelength $\lambda$ is very cpu-intensive.
Accordingly, we have considered only the three grain shapes of 
Fig.\ \ref{fig:shapes}, and only a single size ($a_\eff=0.2\micron$) for
each shape.
\section{Grain Dynamics\label{sec:dynamics}}
\subsection{Equations of Motion}
\begin{figure}
\epsscale{0.5}
\plotone{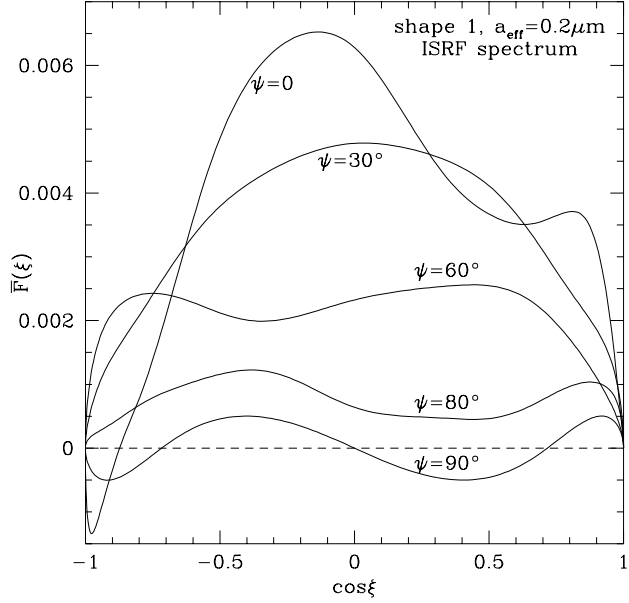}
\caption{
	The precession-averaged function $\bar{F}(\xi)$ affecting
	grain alignment due to radiative torques, where
	$\xi$ is the angle between the grain rotation axis $\ahat_1$ and
	the magnetic field $\bB_0$.
	\label{fig:Fbar.g1}
	}
\end{figure}

\begin{figure}
\epsscale{0.5}
\plotone{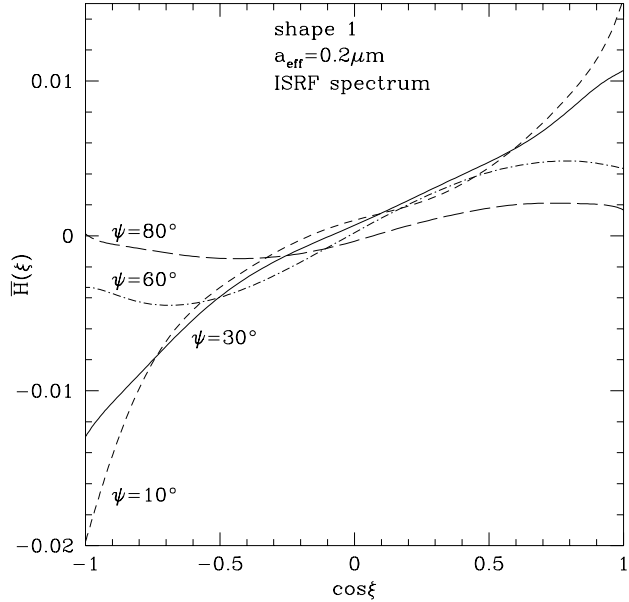}
\caption{
	The precession-averaged function $\bar{H}(\xi)$ 
	for grain with shape 1 and $a_\eff=0.2\micron$.
	$\bar{H}(\xi)$ describes the spinup torque associated with
	anisotropy of the radiation field, where
	$\xi$ is the angle between the grain rotation axis $\ahat_1$ and
	the magnetic field $\bB_0$.
	\label{fig:Hbar.g1}
	}
\end{figure}

From eq.(\ref{eq:eqofmotion}) we obtain three equations of motion:
\begin{eqnarray}
{d\phi\over dt}&=& 
{\gamma u_{rad}\lambdabar a_\eff^2\over 2 I_1 \omega\sin\xi}G(\xi,\phi)
-\Omega_{\rm B} ~~~,
\label{eq:dphidt}
\\
{d\xi\over dt}&=&
{\gamma u_{rad}\lambdabar a_\eff^2\over 2 I_1 \omega}F(\xi,\phi)
-{\sin\xi\cos\xi\over\tau_\DG} ~~~,
\label{eq:dxidt}
\\
{d\omega\over dt}&=&
{u_{rad}\lambdabar a_\eff^2\over 2 I_1}
\left[
	\gamma H(\xi,\phi) + 
	(1-\gamma)\langle Q_\Gamma^\iso\rangle
\right]
-{\omega\sin^2\xi\over\tau_\DG}
+{\omega_{\HH} p(t) \over \tau_{drag}}
-{\omega\over\tau_{drag}} ~~~.
\label{eq:domegadt}
\end{eqnarray}
Because the precession rate $\Omega_{\rm B}$ [see eq.(\ref{eq:omegab})] 
is so much faster than the
other terms in eq.(\ref{eq:dphidt}-\ref{eq:domegadt}) we may immediately
average over $\phi$ in eq.(\ref{eq:dxidt},\ref{eq:domegadt}):
\begin{eqnarray}
{d\xi\over dt^\prime}
&\approx& {\gamma\Mfac\over\omega^\prime}
\bar{F}(\xi) - \beta\sin\xi\cos\xi ~~~,
\label{eq:dxidtp}
\\
{d\omega^\prime\over dt^\prime}
&\approx&
\Mfac
\left[\gamma\bar{H}(\xi) +(1-\gamma)\langle Q_\Gamma^\iso\rangle\right]
+\eta p(t) - \left(1+\beta\sin^2\xi\right)\omega^\prime ~~~,
\label{eq:domegapdtp}
\end{eqnarray}
where $t^\prime\equiv t/\tau_{drag}~$, $~~~\omega^\prime\equiv\omega/\omega_T~$,
and
\begin{eqnarray}
\label{eq:beta}
\beta&\equiv&{\tau_{drag}\over\tau_\DG}~~~,
\\
\eta&\equiv&{\omega_{\HH}\over\omega_T}~~~,
\\
\Mfac &\equiv& 
{u_{rad}\lambdabar a_\eff^2 \tau_{drag}\over 2I_1\omega_T}
={6.83\times10^4\over\delta}
\left({u_{rad}\over \nH kT}\right)
\left({\lambdabar\over \micron}\right)
\left({\alpha_1\rho a_{-5}\over3\g\cm^{-3}}\right)^{1/2} ~~~~,
\label{eq:Mfactor}
\\
\bar{F}(\xi)&\equiv&{1\over2\pi}\int_0^{2\pi}F(\xi,\phi)d\phi
\label{eq:fbar} ~~~~,
~~~~~~~
\bar{H}(\xi)\equiv{1\over2\pi}\int_0^{2\pi}H(\xi,\phi)d\phi ~~~~.
\label{eq:hbar}
\end{eqnarray}
[Note that the parameter $\beta$ in eq.\ (\ref{eq:beta}) is not to be 
confused with the angle $\beta$ in Fig.\ \ref{fig:ThetaPhi}.]
For the numerical examples in this paper, we assume conditions appropriate
to an interstellar diffuse cloud:
$\nH=30\cm^{-3}$, $n(\H)/\nH=1$, $T=100\K$, $B=5\mu\G$.
We consider a grain of silicate composition with 
$\rho=3\g\cm^{-3}$, $K=10^{-13}\s$, and
$a_{-5}=2$.
For most examples we assume $f=1/3$, $l=10\Angstrom$, and $E=0.2\eV$.
Thus
\begin{eqnarray}
\beta &=& 2.87\times10^{-2}\left({2\over\delta}\right) ~~~~,
\label{eq:betaval}
\\
\eta &=& 272 \left({\alpha_1\over 1.5}\right)^{1/2}
\left({2\over\delta}\right)
f \left({l\over10\Angstrom}\right) ~~~~,
\\
\Mfac &=& 1.49\times10^5 \left({\alpha_1\over 1.5}\right)^{1/2}
\left({2\over\delta}\right) ~~~~.
\end{eqnarray}
The functions $\bar{F}(\xi)$ and $\bar{H}(\xi)$ are plotted in Figures
\ref{fig:Fbar.g1} and \ref{fig:Hbar.g1} for shape 1, $a_\eff=0.2\micron$, and
the interstellar radiation field.

\subsection{Stationary Points}
Because $p(t)$ in eq.\ 
(\ref{eq:gamma_h2},\ref{eq:gamma_h2b},\ref{eq:domegadt},\ref{eq:domegapdtp}) 
fluctuates, there are no true steady-state
solutions.
If the surface site lifetime $t_L\gg \tau_{drag}$, 
then there will be quasi-steady
solutions
$\xi_s(p)$, $\omega_s(p)$; $\xi_s$ is a zero of the function
\beq
Z(\xi)\equiv \bar{F}(\xi)- 
{\beta\sin\xi\cos\xi\over1+\beta\sin^2\xi}
\left[\gamma\bar{H}(\xi)+(1-\gamma)\langle Q_\Gamma^\iso\rangle+
{\eta p\over\Mfac}\right]~~~~,
\label{eq:Zfunc}
\eeq
and $\omega_s$ is given by
\beq
\omega_s^\prime = {\omega_s\over\omega_T} = 
{
\Mfac
\left[
	\gamma\bar{H}(\xi_s)+(1-\gamma)\langle Q_\Gamma^\iso\rangle
\right]
+\eta p
\over 
1+\beta\sin^2\xi_s
}~~~~.
\eeq

From Figs.\ \ref{fig:Fbar.g1} and \ref{fig:Hbar.g1} we see that $\bar{F}$
and $\bar{H}$ are of comparable magnitude.
Since for conditions of interest we have $\gamma\approx0.1$,
$M\gg1$, and $\beta\ll 1$ (see eq.\ \ref{eq:Mfactor},\ref{eq:betaval}),
it is apparent from eq.\ (\ref{eq:Zfunc}) that $Z(\xi)\approx\bar{F}(\xi)$,
so that the zeros of $Z$ will nearly coincide with the zeros of $\bar{F}$.
The $\HH$ torques and Davis-Greenstein torques have only a minimal
effect on the function $Z(\xi)$, and hence on  the angles
$\xi_s$ for stationary solutions
(although the angular velocities $\omega_s$ are significantly dependent
on the $\HH$ torques, as will be seen below).

The timescale for approach to alignment angle $\xi_s$ 
may be estimated by linearizing eq.\
(\ref{eq:dxidtp},\ref{eq:domegapdtp}) around $(\xi_s,\omega_s)$:
\begin{eqnarray}
{d\xi\over dt^\prime}
&\approx&
A(\xi-\xi_s) +B(\omega^\prime-\omega_s^\prime) ~~~,
\\
{d\omega^\prime\over dt^\prime}
&\approx&
C(\xi-\xi_s) +D(\omega^\prime-\omega_s^\prime) ~~~,
\end{eqnarray}
where
\begin{eqnarray}
A&=&
{\gamma\Mfac\over\omega_s^\prime}{d\bar{F}\over d\xi}-
\beta\cos2\xi_s ~~~,
\\
B&=&
- 
{\gamma\Mfac\over(\omega_s^\prime)^2}\bar{F}(\xi_s) ~~~,
\\
C&=&
\gamma\Mfac{d\bar{H}\over d\xi} - \beta\omega_s^\prime\sin2\xi_s ~~~,
\\
D&=&
-
\left(1+\beta\sin^2\xi_s\right) ~~~.
\end{eqnarray}
The stationary point $(\xi_s,\omega_s)$ is an attractor
(i.e., stable) provided both
\beq
A+D<0 ~~~~~{\rm and}~~~~~BC-AD<0 ~~~;
\eeq
otherwise it is a repellor (i.e., unstable).
If stable, there will be two relaxation times for approach to the
stationary point, given by
\beq
\tau_{relax}^{-1} 
=
{-(A+D)\pm\left[(A-D)^2+4BC\right]^{1/2}\over 2\tau_{drag}} ~~~.
\eeq
We take the slower of these two relaxation rates to give the
alignment time:
\beq
\tau_{align}^{-1}
=
{-(A+D)-\left[(A-D)^2+4BC\right]^{1/2}\over 2\tau_{drag}} ~~~.
\label{eq:talign}
\eeq

\subsection{Crossover Points}

When $\omega\rightarrow0$, eq.(\ref{eq:dxidtp}) is singular unless
$\bar{F}=0$.
Therefore $\omega$ can only change sign at a crossover point $\xi_c$
where 
\beq
\bar{F}(\xi_c)=0 ~~~~.
\label{eq:xic}
\eeq
From eq.(\ref{eq:F(xi,phi)}) and eq.(\ref{eq:fbar}) it is easily seen
that $\xi_c=0,\pi$ are always solutions to eq.(\ref{eq:xic}).
The crossover point acts as an attractor if
\beq
{\gamma M  \over
M[\gamma \bar{H}(\xi_c)+(1-\gamma)\langle Q_\Gamma^\iso\rangle]+\eta p}
\left({d\bar{F}\over d\xi}\right) > 0 ~~~~,
\label{eq:xattractor}
\eeq
and as a repellor if eq.(\ref{eq:xattractor}) is not satisfied.
Physical ``crossovers'' will only occur at crossover points which are
attractors.

We define the ``polarity'' of a crossover point as positive or negative
according to the sign of $d\omega/dt$.
From eq.(\ref{eq:domegapdtp}) we see that
\beq
{\rm polarity} ~= {\rm sign}
\left\{
	M[\gamma \bar{H}(\xi_c)+
	(1-\gamma)\langle Q_\Gamma^\iso\rangle]+\eta p
\right\} ~~~~.
\label{eq:polarity}
\eeq

\section{Trajectory Maps\label{sec:traj_maps}}
\subsection{Map Classification}

Provided the rotation is highly superthermal (so that we may assume that
$\ahat_1\parallel\bJ/\omega$), and rotation and precession are both rapid compared
to the timescale for changes in $\omega$ or $\xi$,
the grain state is specified by two coordinates: the angle $\xi$ between
$\bJ$ and $\bB_0$, and the rate $\omega$ of rotation around $\ahat_1$.
The dynamical evolution of the grain therefore corresponds to a trajectory
on the $\xi,\omega$ plane.
Each point on the $\xi,\omega$ plane lies on a single trajectory, except for
(1) stationary attractors where many different trajectories have a common
end point, or (2) crossover attractors, where many
different trajectories converge.

We refer to the set of all trajectories as the ``trajectory map''.
The $\xi,\omega$ plane naturally separates into two
regions, $\omega>0$ and $\omega<0$, with trajectories connecting the
regions only at a small number of crossover attractors (a crossover repellor
formally admits a single trajectory connecting the two regions but this
is a set of zero area on the $\xi,\omega$ plane).
To a considerable extent it is possible to classify trajectory maps
according to the types of stationary points and crossover points
which they possess.
Three distinct types of trajectory maps are possible.

\subsubsection{Cyclic Maps}

If a trajectory map has no stationary attractors, then the generic trajectory
does not end.
Such trajectory maps are referred to as ``cyclic''.
Fig.\ \ref{fig:tr_30.g1p} is an example of a cyclic map.

From eq.(\ref{eq:dxidtp}) we see that if $\beta=0$ (i.e.,
$\tau_\DG\rightarrow\infty$: no Davis-Greenstein torque),
then the sign of $d\xi/dt$ is independent of $\omega$ (except for the
sign of $\omega$).
Since a closed trajectory must have two points with the same $\xi$ but
opposite signs of $d\xi/dt$, it follows that there can be 
no closed trajectories contained within either the
$\omega>0$ or $\omega<0$ regions.
If closed trajectories exist, then each such trajectory
must pass through $\omega=0$ at an even number of crossover attractors, and
the trajectory must have equal numbers of crossover
attractors of positive and negative polarity (usually just one of each).

For finite $\tau_\DG$ closed trajectories which do not cross $\omega=0$ 
could, in principle, exist,
depending on the properties of $\bar{F}(\xi)$ and $\bar{H}(\xi)$
(although any such trajectories cannot cross $\xi=0$).\footnote{
	If a closed trajectory crossed $\xi=0$, then there would be
	at least two different points on the trajectory with
	$\xi=0$ but opposite signs for $d\xi/dt$.
	It is clear from eq.(\ref{eq:dxidtp}) that this
	is not possible since the Davis-Greenstein torque term
	vanishes at $\xi=0$.
	}
However, since $\beta\ll1$ for interstellar conditions of interest,
it seems safe to assume that closed trajectories do not occur unless they
cross $\omega=0$.
This will henceforth be assumed.

\subsubsection{Noncyclic Maps}

If the trajectory map does not have at least one crossover attractor
of each parity
then,
following the discussion above, there will be no closed trajectories.
We refer to this type of map as ``noncyclic''.

\begin{figure}
\plotone{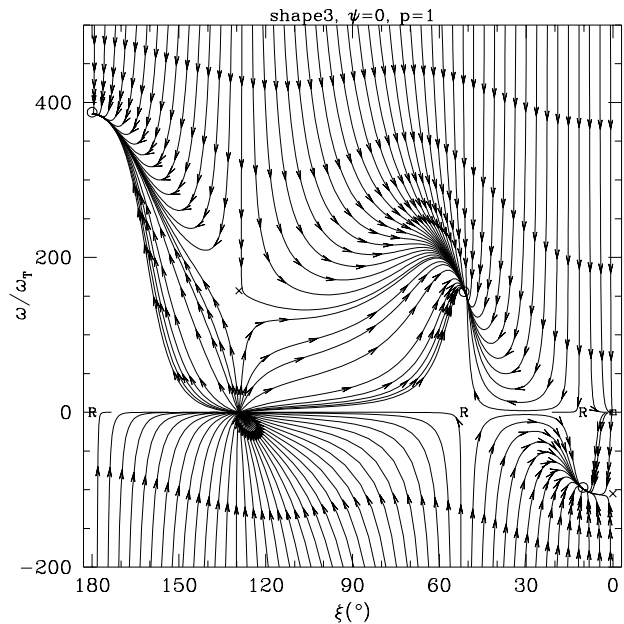}
\caption{
	\label{fig:tr_00.g3p}
	Trajectory map for shape 3 and $\psi=0$, an example of
	a map which has crossover attractors of both polarities, but
	with no closed trajectories: all trajectories terminate on one
	of the three stationary attractors.
	The three crossover repellors are at the points labelled by R.
	}
\end{figure}

In a noncyclic map,
every trajectory must terminate at a stationary attractor point,
on a timescale given by eq.\ (\ref{eq:talign});
thus there must be at least one stationary attractor.
Figs.\ \ref{fig:tr_60.g1p} -- \ref{fig:tr_00.g3p} are examples of
noncyclic trajectory maps.

\subsubsection{Semicyclic Maps}

The case where there are both closed trajectories but also at least one
stationary attractor will be referred to as ``semicyclic''.
A necessary condition for a semicyclic map is that there be at least
one crossover
attractor of each polarity, and at least one stationary
attractor.
If there is no crossover repellor situated between two crossover
attractors of opposite polarity, then there will be trajectories 
with $|\omega|\rightarrow0$
connecting the two crossover attractors.
Therefore a {\it sufficient} condition for a semicyclic map is that there
be at least one stationary attractor, and at least one pair
of crossover attractors of opposite parity with no crossover repellor
in between.
Figs.\ \ref{fig:tr_00.g1z} and \ref{fig:tr_00.g1p} are examples of
semicyclic maps.

If there is a repellor situated between each pair of
opposite-polarity crossover attractors, there may be no closed trajectories:
Fig.\ \ref{fig:tr_00.g3p} is an example of such a noncyclic trajectory map.
We conjecture that whenever a crossover repellor lies between each
pair of opposite-polarity crossover attractors, the trajectory map will
be noncyclic.

\subsection{Examples}

\begin{figure}
\epsscale{0.80}
\plotone{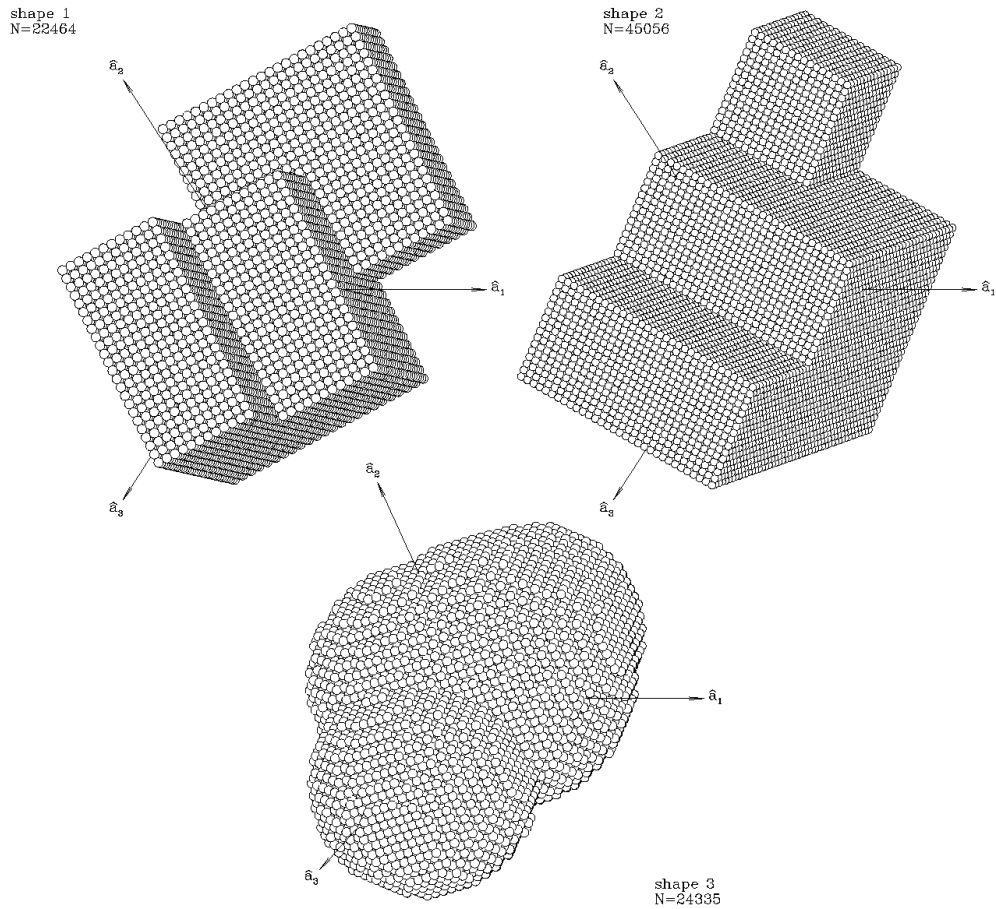}
\caption{
	Three representative grain geometries.  Shape 1 is the geometry
	studied in Paper I.
	\label{fig:shapes}
	}
\end{figure}

\begin{figure}
\epsscale{0.5}
\plotone{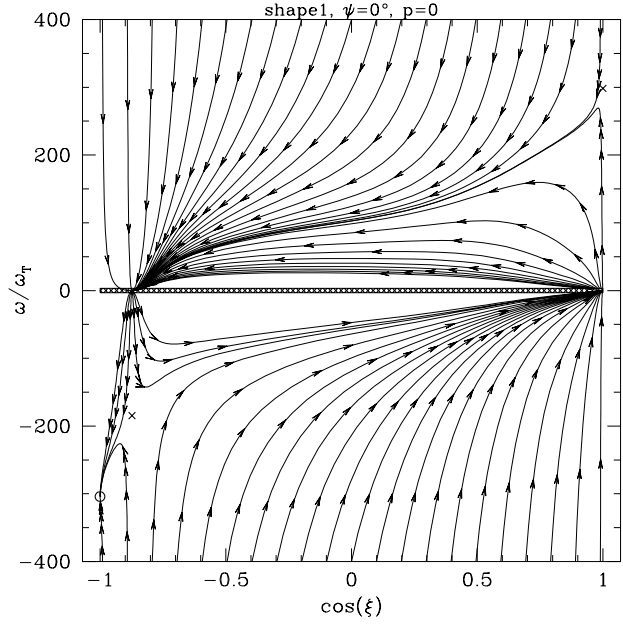}
\caption{
	\label{fig:tr_00.g1z}
	Trajectories on the $\cos\xi$, $\omega$ plane, for $\psi=0$,
	with only radiative torques and gas drag acting.
	The grain has the geometry of shape 1 in 
	Fig.\ \protect{\ref{fig:shapes}}
	and an effective radius $a_\eff=0.2\micron$.
	The gas is assumed to have $\nH=30\cm^{-3}$ and $T=100\K$.
	The radiation field is assumed to have $u_{rad}/\nH kT=2.09$,
	and an anisotropy factor $\gamma=0.1$.
	Arrows are at intervals of $0.5t_{drag}=7.5\times10^4\yr$.
	There are three stationary points: 
	two repellors (crosses) and one attractor (circle).
	There are two ``attractor'' crossover points at $\cos\xi=-.873$ and 1
	(the crossover point at $\cos\xi=-1$ is a repellor).
	As discussed in the text, the analysis is uncertain in the region 
	$(\omega/\omega_T)^2<10$, which has been shaded.
	}
\end{figure}

In Figure \ref{fig:tr_00.g1z} we show evolutionary trajectories for 
an $a_\eff=0.2\micron$ silicate grain with the geometry of shape 1 in
Fig.\ \ref{fig:shapes}, for $\psi=0$
(anisotropic component of the radiation field parallel to $\bB_0$).
We show results assuming $p=0$ (no $\HH$ torques) and $\tau_\DG=\infty$
(no Davis-Greenstein torque): the only torques are due to gas drag and
starlight.
This is an example of a semicyclic map.
The arrows in the figure are at intervals of $0.5t_{drag}=7.5\times10^4\yr$.
We see that trajectories beginning at $\omega > 0$ all converge on
the crossover point at $\cos\xi=-.873$, 
where they enter the region $\omega<0$.
Some of the trajectories in the region $\omega<0$ evolve to the stationary
attractor at $\cos\xi=-1$; others lead to the crossover attractor at
$\cos\xi=1$, where they enter the $\omega>0$ region.

\begin{figure}
\epsscale{0.5}
\plotone{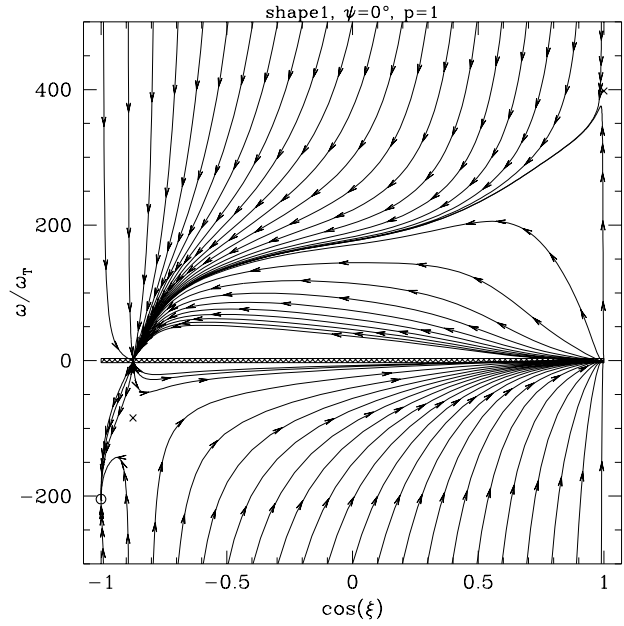}
\caption{
	\label{fig:tr_00.g1p}
	Same as Fig.\protect{\ref{fig:tr_00.g1z}}, but now including torques
	due to
	$\HH$ formation (with $f=1/3$, $l=10\Angstrom$,
	$p=1$, and $n(\H)/\nH=1$),
	and paramagnetic dissipation with
	$K=10^{-13}\s$, and $B=5\mu\G$.
	There are three stationary points: 
	one attractor (circle) and two repellors (crosses).
	}
\end{figure}

\begin{figure}
\epsscale{0.5}
\plotone{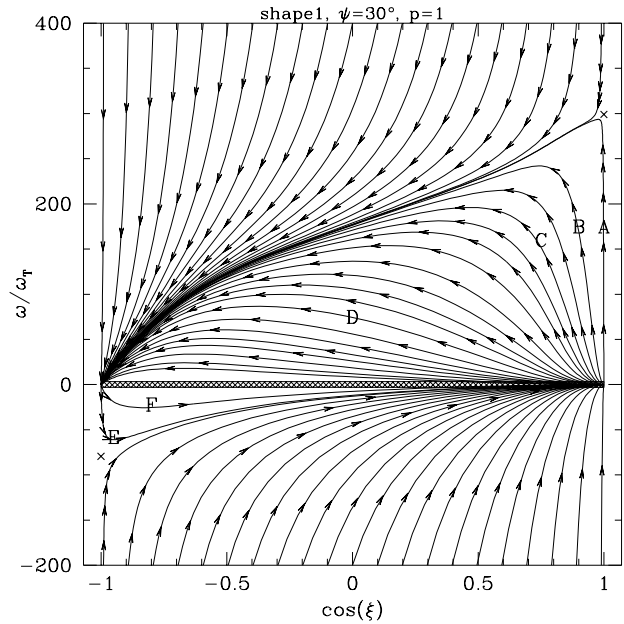}
\caption{
	\label{fig:tr_30.g1p}
	Same as Fig. \protect\ref{fig:tr_00.g1p},
	but for $\psi=30\arcdeg$.
	For this case there are no attractors, and two repellors.
	The grain trajectory presumably ends up cycling around and
	around, crossing the $\omega=0$ axis at the ``crossover''
	points at $\xi=0$ and $180\arcdeg$, both of which are ``attractors''.
	Labels A--F designate trajectories examined in 
	Fig.\protect{\ref{fig:dpdcosxi}}.
	}
\end{figure}

\begin{figure}
\epsscale{0.5}
\plotone{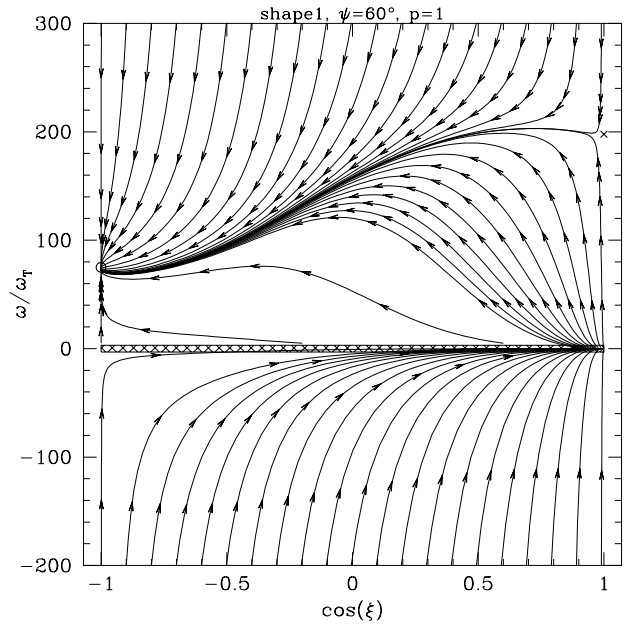}
\caption{
	\label{fig:tr_60.g1p}
	Same as Fig. \protect\ref{fig:tr_00.g1p},
	but for $\psi=60\arcdeg$.
	For this case there is one attractor and one repellor.
	}
\end{figure}

\begin{figure}
\epsscale{0.5}
\plotone{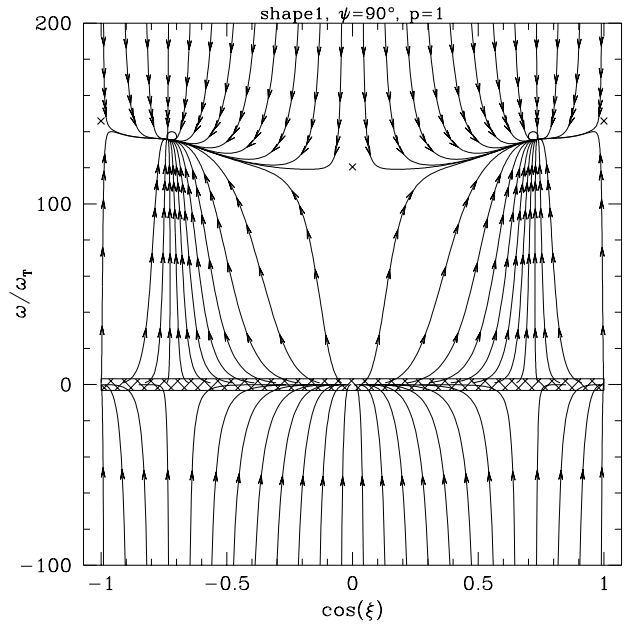}
\caption{
	\label{fig:tr_90.g1p}
	Same as Fig. \protect\ref{fig:tr_00.g1p},
	but for $\psi=90\arcdeg$.
	For this case there are two attractors (at $\cos\xi=\pm0.71$ 
	and three repellors.
	}
\end{figure}

As mentioned above, in our analysis the crossover points are singular --
we are unable to 
follow a grain's evolution through
the crossover event, and therefore we cannot predict which of the trajectories
emerging from a crossover attractor will be ``populated''.
Furthermore, our analysis assumes superthermal rotation with 
perfect Barnett relaxation
so that $\ahat_1\parallel\bJ/\omega$.
This approximation will be accurate when the grain rotation is extremely
superthermal, $(\omega/\omega_T)^2\gg1$, but fails when the rotational
kinetic energy approaches thermal values.
The region $(\omega/\omega_T)^2<10$ is shaded in Figs.\ 
\ref{fig:tr_00.g1z}-\ref{fig:tr_90.g1p};
with our present approximations we are unable to follow trajectories within
this region.

If we now include $\HH$ formation torques (with $p=+1$) and
Davis-Greenstein torques due to normal paramagnetism 
($\tau_\DG=1.0\times10^7\yr$) the ``flow pattern'' changes to that shown
in Figure \ref{fig:tr_00.g1p}.
The map remains semicyclic.
Figures \ref{fig:tr_00.g1z} and \ref{fig:tr_00.g1p} are qualitatively similar,
showing that $\HH$ torques, while by no means negligible, appear to be
of secondary importance compared to radiative torques for $a_\eff=0.2\micron$
grains.

Figures \ref{fig:tr_00.g1z} and \ref{fig:tr_00.g1p} each show 
semicyclic behavior.
Some trajectories (e.g., from $\omega=-300\omega_T$ and $\cos\xi=-.9$)
proceed directly to a stable attractor where they are captured.
Other trajectories (e.g., from $\omega=-300\omega_T$, $\cos\xi=-.8$)
head for the crossover attractor.
In Figures \ref{fig:tr_00.g1z} and \ref{fig:tr_00.g1p} 
it is clear that even though we are unable to
predict how the grain will emerge from the crossover point at $\cos\xi=1$,
we know that the grain must then proceed to the other crossover attractor
at $\cos\xi=-.88$.
At this crossover the details matter: some trajectories emerging from the
crossover lead to the stable attractor, while other trajectories return to
the crossover point at $\cos\xi=+1$.
In principle, it is possible that a grain could cycle back and forth
between these two crossover points many times; it then becomes 
essential to understand the mechanics (and statistics) of the crossover
process.
Such cyclic behavior will in general be a possibility whenever we have
two crossover attractors of opposite polarity with no crossover repellor
in between.

Figure \ref{fig:tr_30.g1p} shows the trajectory map for the same grain as in 
Figure \ref{fig:tr_00.g1p}, but now for an angle $\psi=30\arcdeg$ 
between $\bB_0$ and the starlight anisotropy.
We now have two stationary points rather than three, and there are no
stationary attractors: the grain must continuously cycle between the two
crossover attractors, which again have opposite polarity.
This is a clear example of a ``cyclic'' trajectory map.

If we rotate the anisotropy direction to $\psi=60\arcdeg$
(Fig.\ \ref{fig:tr_60.g1p}) the stationary
point at $\cos\xi=-1$ (which for $\psi=30\arcdeg$ was a repellor) 
crosses over into the $\omega>0$ region and becomes
a stable attractor.
The trajectory map now has a straightforward structure:
there is only one stable attractor and only one stable crossover.
The grain {\it must} evolve to the $\omega>0$ region, 
where it is inevitably trapped by the stable attractor.

The case $\psi=90\arcdeg$ 
(Fig.\ \ref{fig:tr_90.g1p}) shows new possibilities.  
Because of the symmetry of the problem, this trajectory map is symmetric
under reflection through $\cos\xi=0$.
There are five stationary points: two stationary attractors plus three
stationary repellors.
There are three crossover attractors, at $\cos\xi=\pm1$ and 0; all have
positive polarity.
All trajectories lead to a stable attractor with
$\cos^2\xi=0.514$.
This is another example of a noncyclic trajectory map.

Finally, to illustrate the diversity of possible behaviors, 
Fig.\ \ref{fig:tr_00.g3p} shows the trajectory map for an $a_\eff=0.2\micron$
silicate grain with the geometry of shape 3 (see Fig.\ \ref{fig:shapes})
for an angle $\psi=0$ between the starlight anisotropy direction
and $\bB_0$.
For this case there are four stationary points, three of which are
attractors.
There are two crossover attractors and three crossover repellors.
This is a noncyclic map, but which of the three stationary attractors the
grain is captured by depends on the initial conditions and (for trajectories
which pass through the crossover attractor at $\xi=130\arcdeg$) on the
details of the crossover process, which will determine 
which trajectories emerging from the crossover attractor
are populated.

\section{Discussion\label{sec:discuss}}

\subsection{Grain Alignment}
It is by now evident that radiative torques lead to complex grain dynamics,
which depend delicately on the grain geometry, the angle $\psi$ between
the starlight anisotropy and the magnetic field $\bB_0$, and on the
$\HH$ formation torque to which the grain is subject.

In the Rayleigh limit, the linear dichroism of a population of identical
grains, each spinning around its axis $\ahat_1$, is proportional to
(see, e.g., Lee \& Draine 1985)
\beq
\Delta\kappa =
\langle R\rangle \sin^2\theta_B ~n_{gr}\pi a_{\rm eff}^2 Q_{pol}(\lambda)
\label{eq:dichroism}
\eeq
where $\theta_B$ is the angle between $\bB_0$ and the line-of-sight,
$\langle R\rangle$ is the ensemble-averaged
alignment parameter (or ``Rayleigh reduction factor'')
\beq
R(\xi) \equiv {3\over2}\left(\cos^2\xi-{1\over3}\right)
\eeq
and the polarization efficiency factor
\begin{eqnarray}
Q_{pol}(\lambda) &\equiv&
{1\over2}
\left[ 
Q_{ext}(\bE\parallel\ahat_2,\khat\parallel\ahat_3)+
Q_{ext}(\bE\parallel\ahat_3,\khat\parallel\ahat_2)-\right. \nonumber \\ 
&&
\left. 
~~~~
Q_{ext}(\bE\parallel\ahat_1,\khat\parallel\ahat_2)-
Q_{ext}(\bE\parallel\ahat_1,\khat\parallel\ahat_3)
\right] ~~~,
\end{eqnarray}
where $Q_{ext}$ is the extinction efficiency factor for the grain.
While eq.\ (\ref{eq:dichroism}) for the linear dichroism is strictly
valid only in the Rayleigh limit (where, of course, $Q_{ext}$ depends only
on the direction of $\bE$, but not on $\khat$) 
it provides a good approximation even when the grain is not small compared
to the wavelength.
Hence $\langle R\rangle$ is a useful statistic to characterize the
effectiveness of grain alignment.

\begin{figure}
\epsscale{0.5}
\plottwo{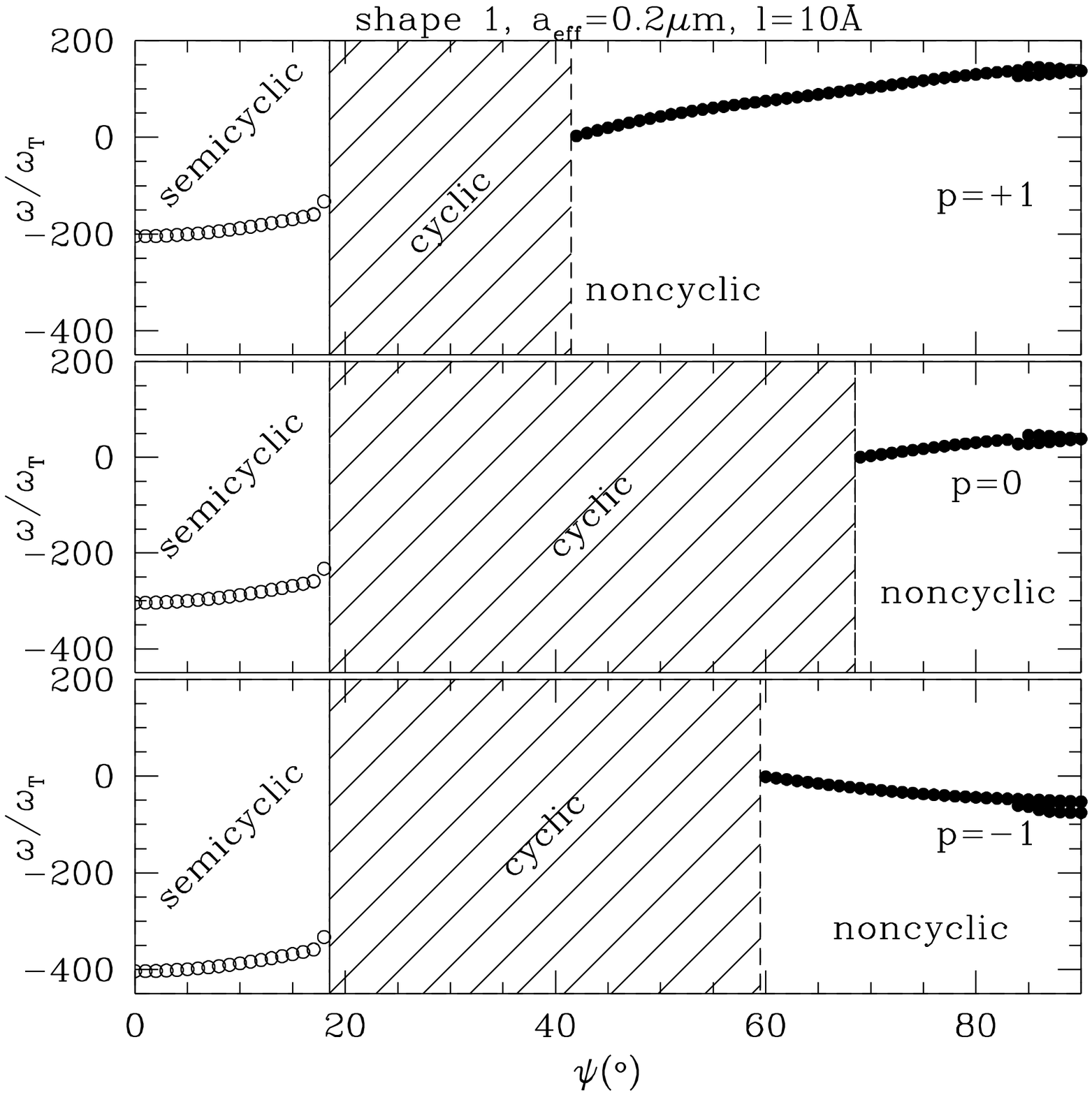}{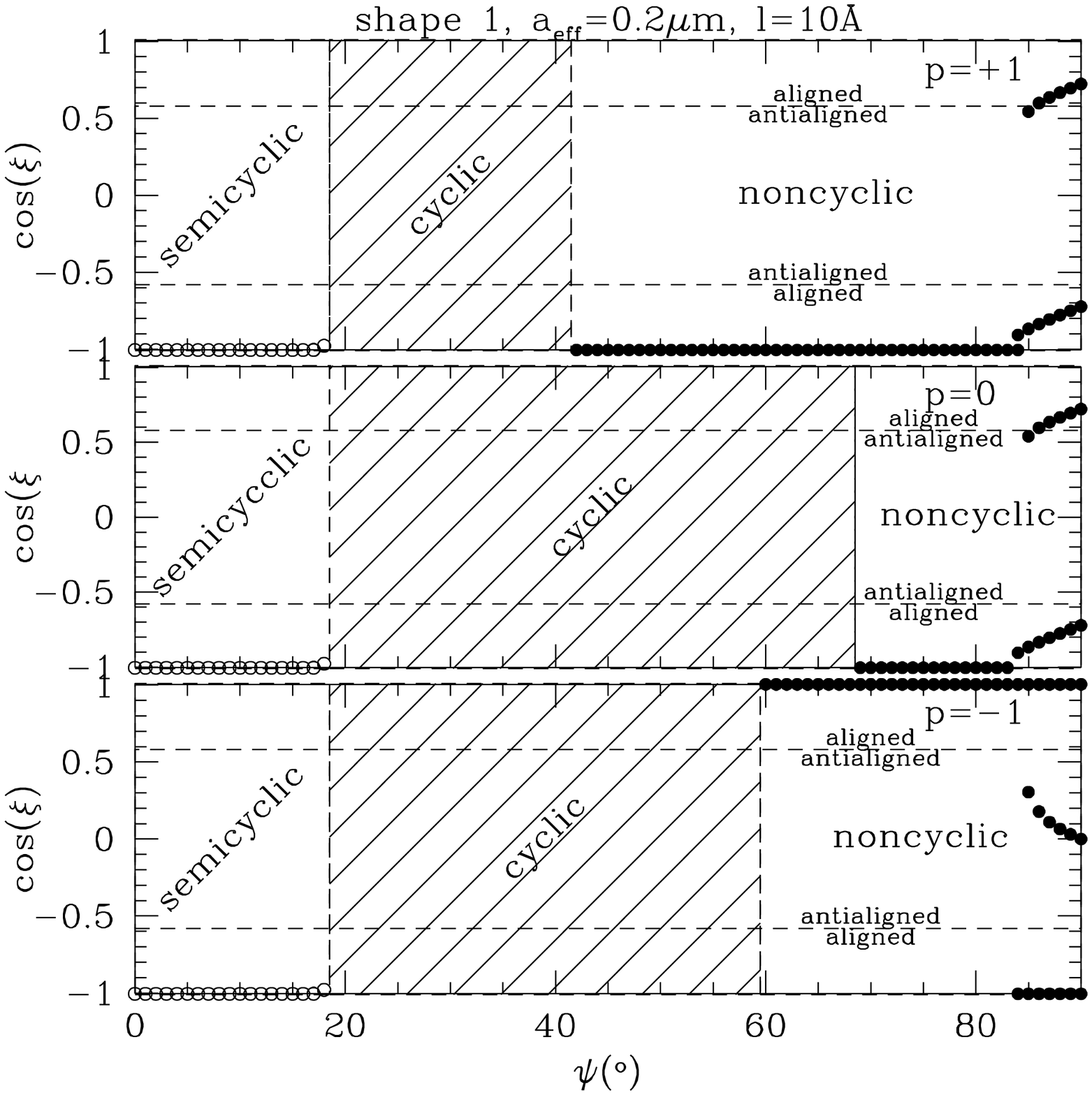}
\caption{
	\label{fig:omega.g1}
	$\omega/\omega_T$ and
	$\cos\xi$
	at attractor stationary points for an $a_\eff=0.2\micron$ silicate
	grain
	with shape 1 in a diffuse cloud, as a function of the angle $\psi$
	between the radiation anisotropy and the magnetic field $\bB_0$.
	A starlight anisotropy $\gamma=0.1$ is assumed.
	Results are shown for $\HH$ torques characterized by 
	$f=1/3$, $l=10\Angstrom$, and $p=+1,0,-1$
	[see eq. (\protect\ref{eq:gamma_h2})].
	In the shaded region there are no attractors, and all trajectories
	are ``cyclic''.
	Open symbols indicate that the trajectory map is ``semicyclic''
	(see text).
	solutions with $\cos\xi>0.577$ or $<-0.577$ are ``aligned'',
	with $R>0$; 
	solutions with $-0.577<\cos\xi<0.577$ are ``antialigned'',
	with $R<0$.
	}
\end{figure}

\begin{figure}
\epsscale{0.5}
\plottwo{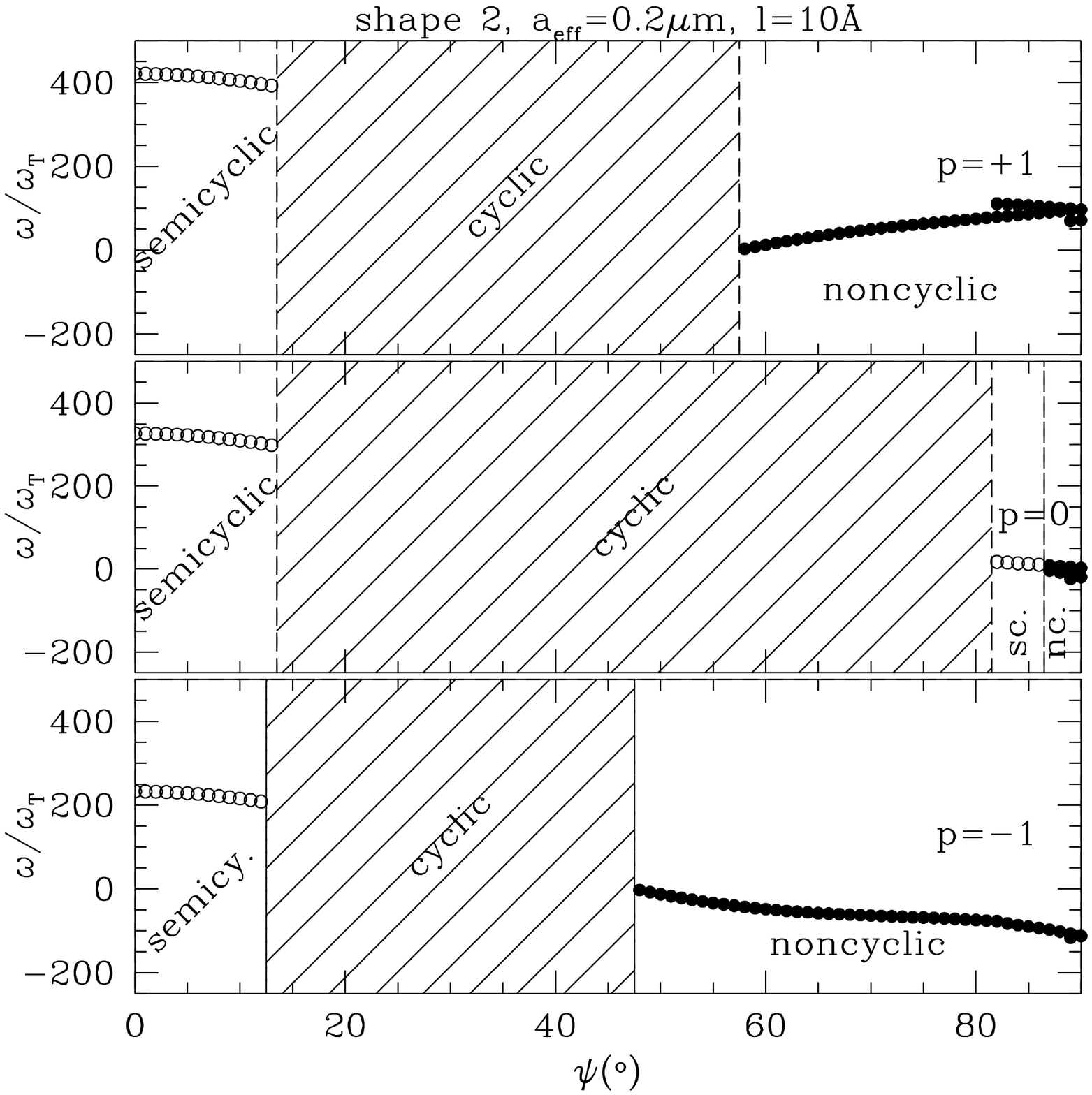}{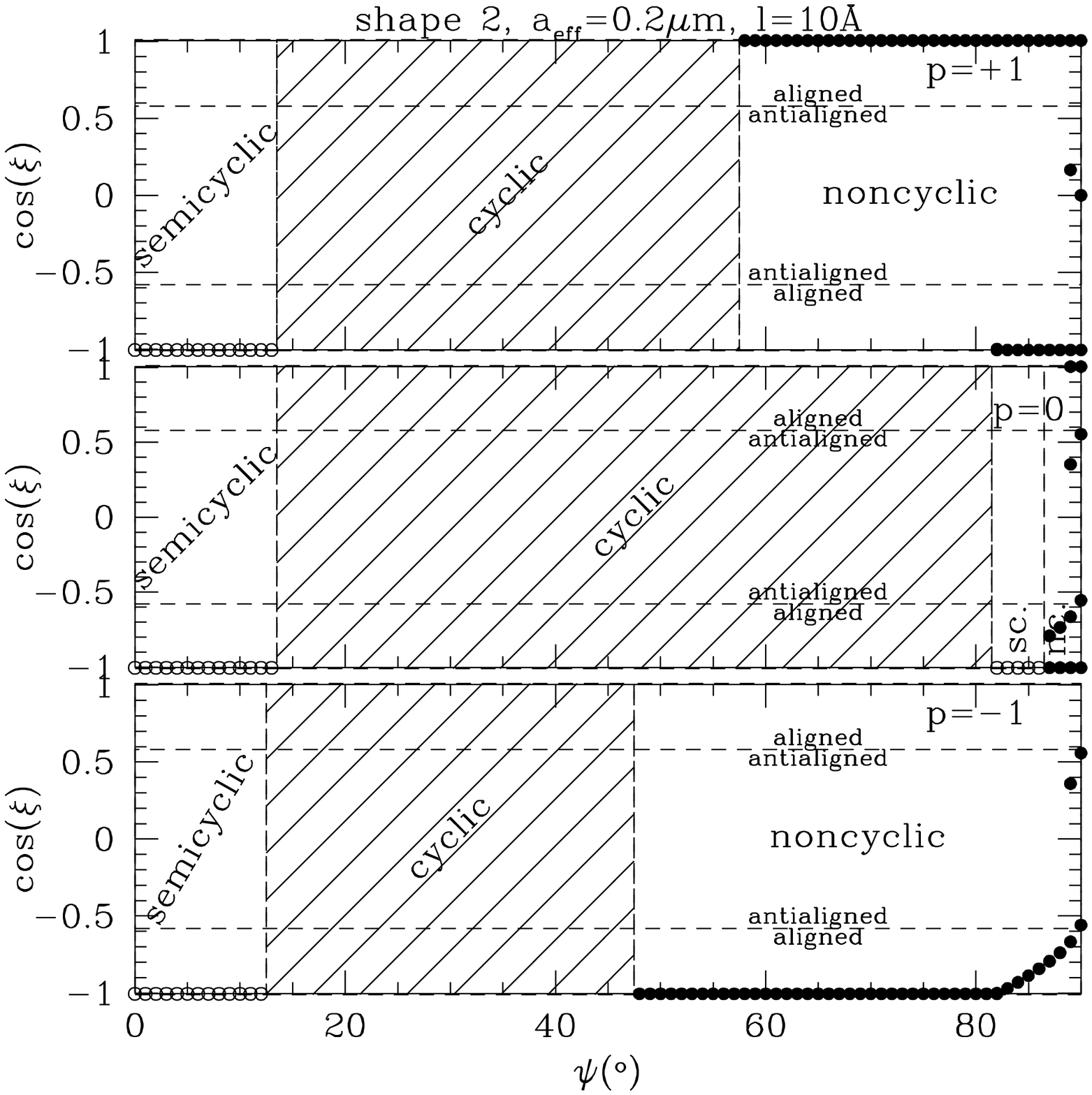}
\caption{
	\label{fig:omega.g2}
	Same as Fig. \protect\ref{fig:omega.g1} but for shape 2.
	}
\end{figure}

\begin{figure}
\epsscale{0.5}
\plottwo{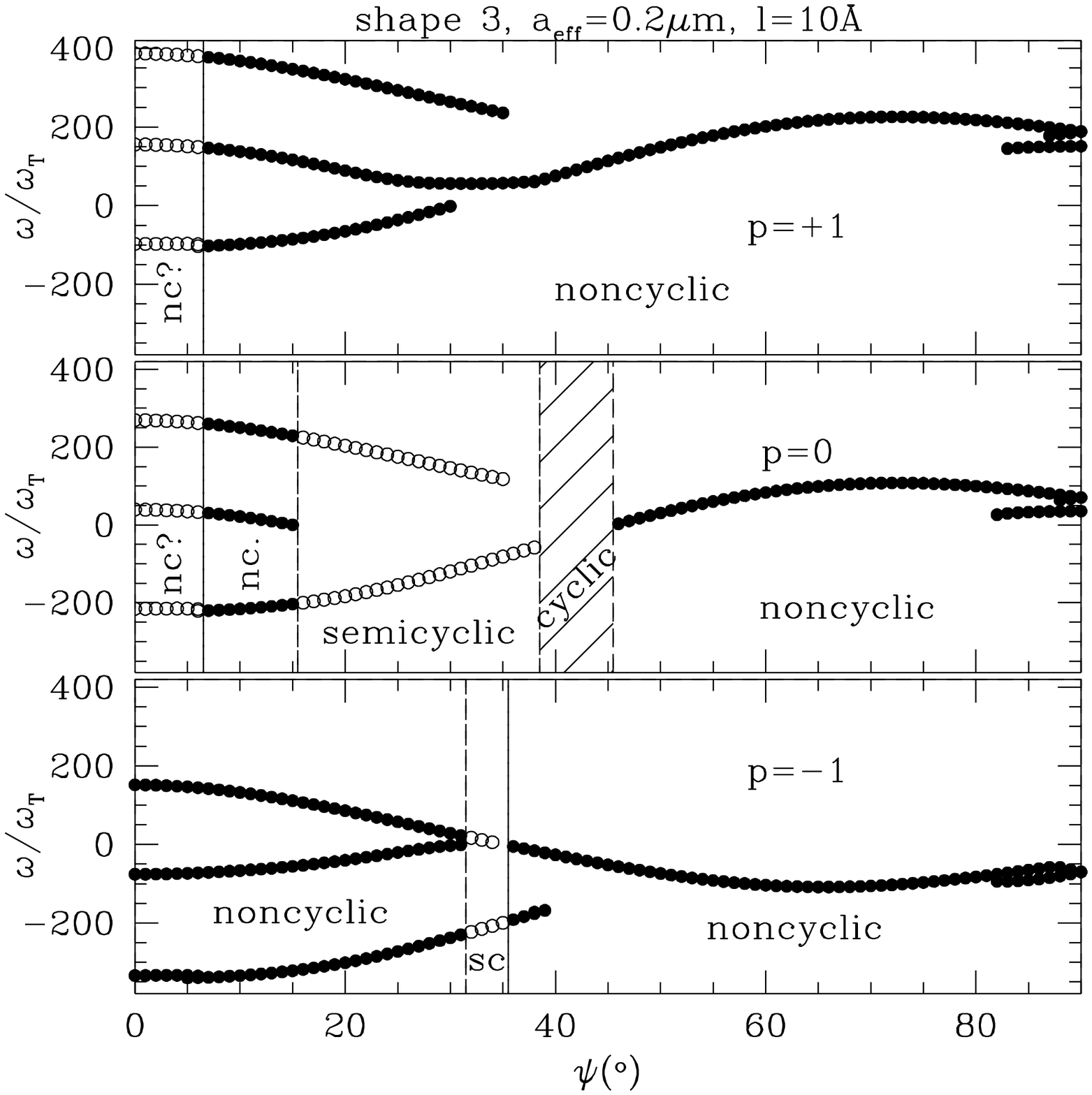}{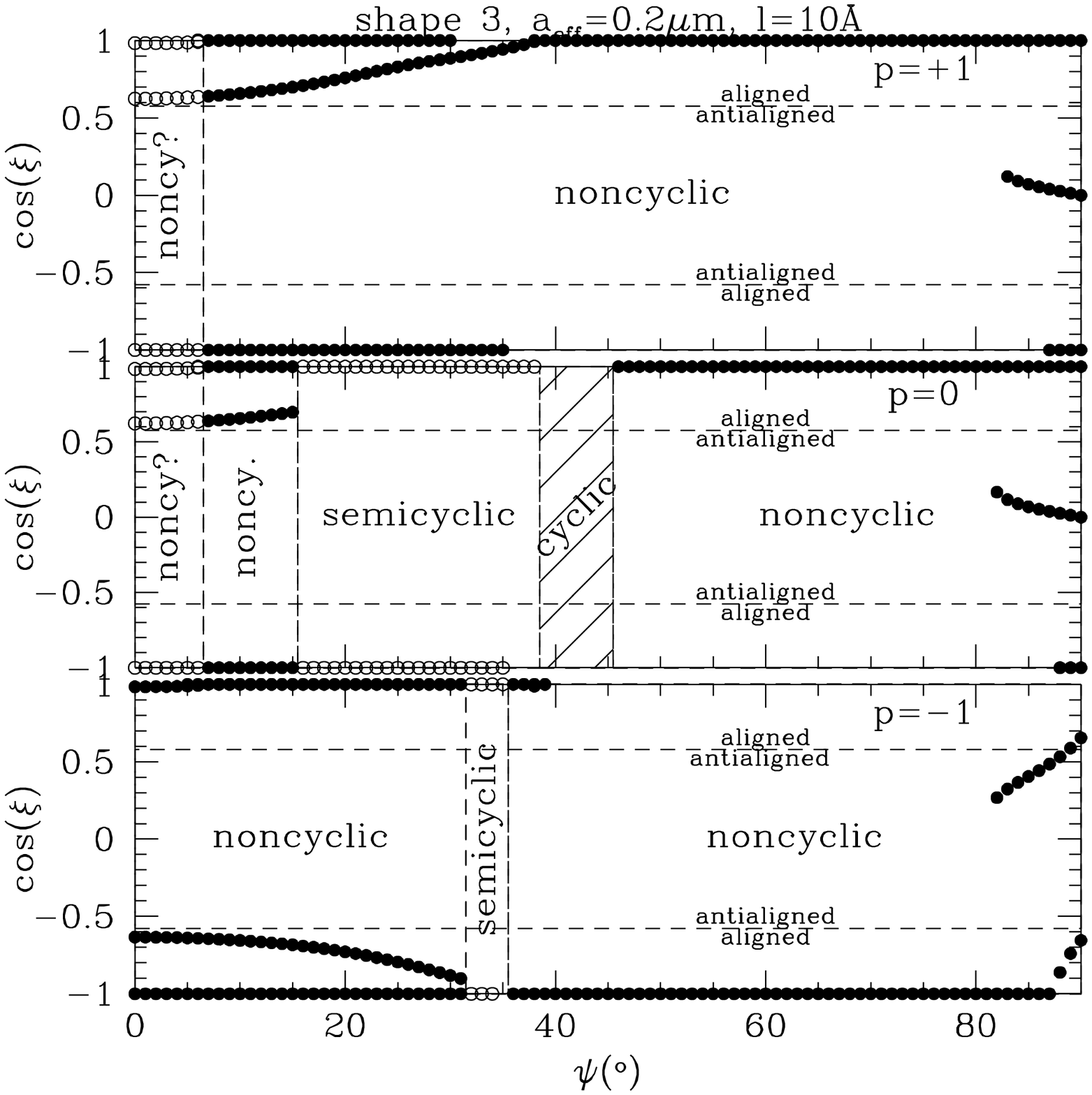}
\caption{
	\label{fig:omega.g3}
	Same as Fig. \protect\ref{fig:omega.g1} but for shape 3.
	}
\end{figure}

Above we have shown 6 examples of trajectory maps, but these are obviously
a very incomplete sampling of parameter space.
To provide a more global picture of the grain dynamics, Figures
\ref{fig:omega.g1}--\ref{fig:omega.g3} show the ``terminal'' values of
the grain rotation rate $\omega$ and $\cos\xi$.
The domains of noncylic, cyclic, and semicyclic solutions are indicated.
The region labelled ``noncyclic?'' is where there is at least one
stationary attractor, and
there are crossover attractors of opposite polarity, but where each 
opposite polarity pair of crossover attractors 
has an intervening crossover repellor.
As discussed above, we suspect that such cases are noncyclic, but we
have not verified this in all cases.

In Figs.\ \ref{fig:omega.g1}--\ref{fig:omega.g3} the horizontal broken lines
show the values of
$\cos\xi=\pm1/\surd3=\pm0.577$ corresponding to alignment parameter $R=0$.
Solutions with $-0.577<\cos\xi<0.577$ are ``antialigned'',
with $R<0$, but it is apparent that antialigned stationary solutions are
relatively rare.
For example, Fig.\ \ref{fig:omega.g1} shows that for $p=1$ and $p=0$,
shape 1 is antialigned only for $\psi=84^\circ$; 
even in this case the antialignment is minimal and there are 
competing stationary solutions with substantial alignment.

From visual inspection of 
Figures \ref{fig:omega.g1}--\ref{fig:omega.g3}, we conclude that
in a substantial fraction of cases, the grains have noncyclic trajectory
maps with stationary solutions with appreciable, and often perfect,
alignment.

\subsection{Sensitivity to the Magnitude of the H$_2$ Formation Torque}

\begin{figure}
\epsscale{0.5}
\plottwo{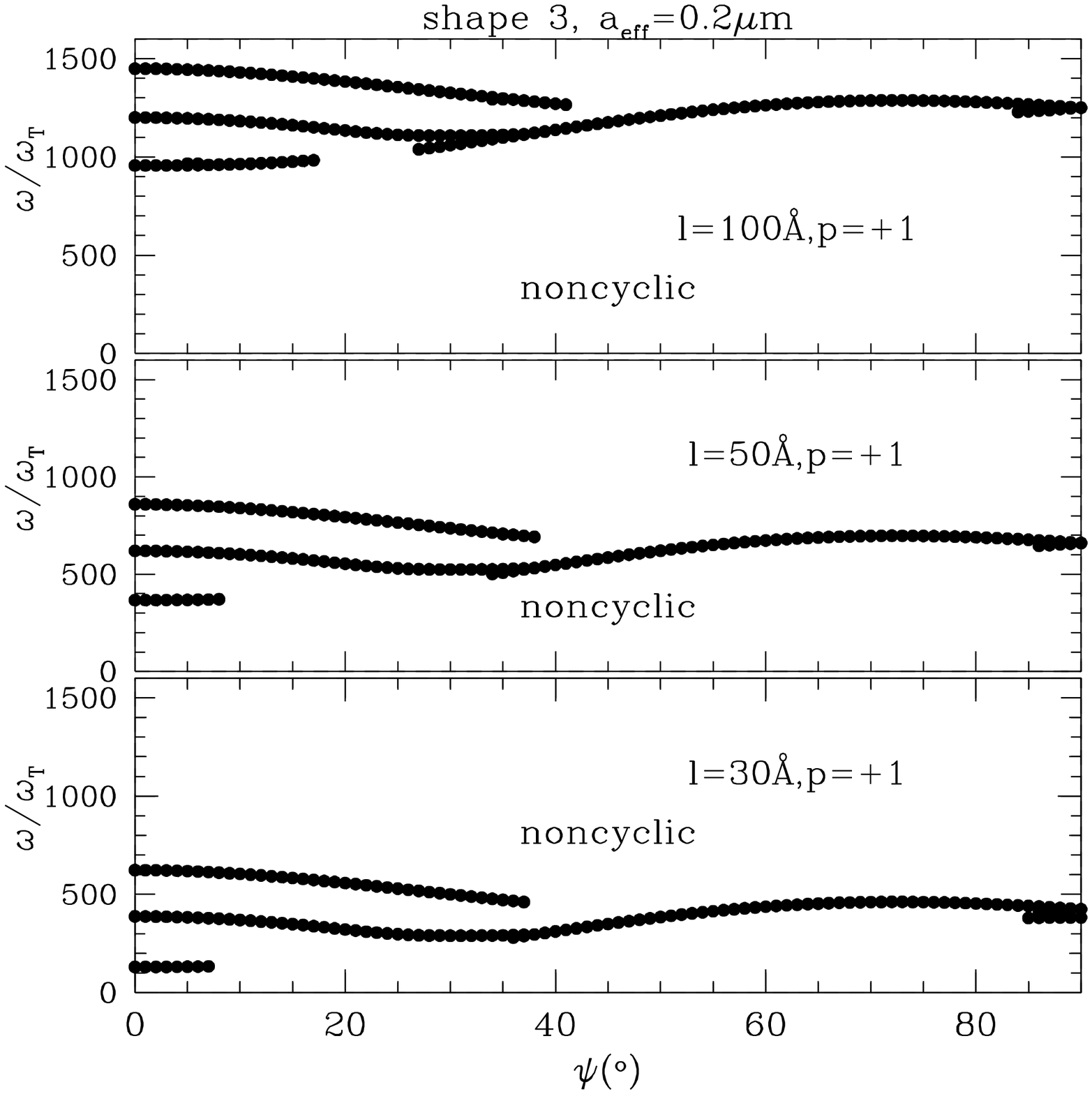}{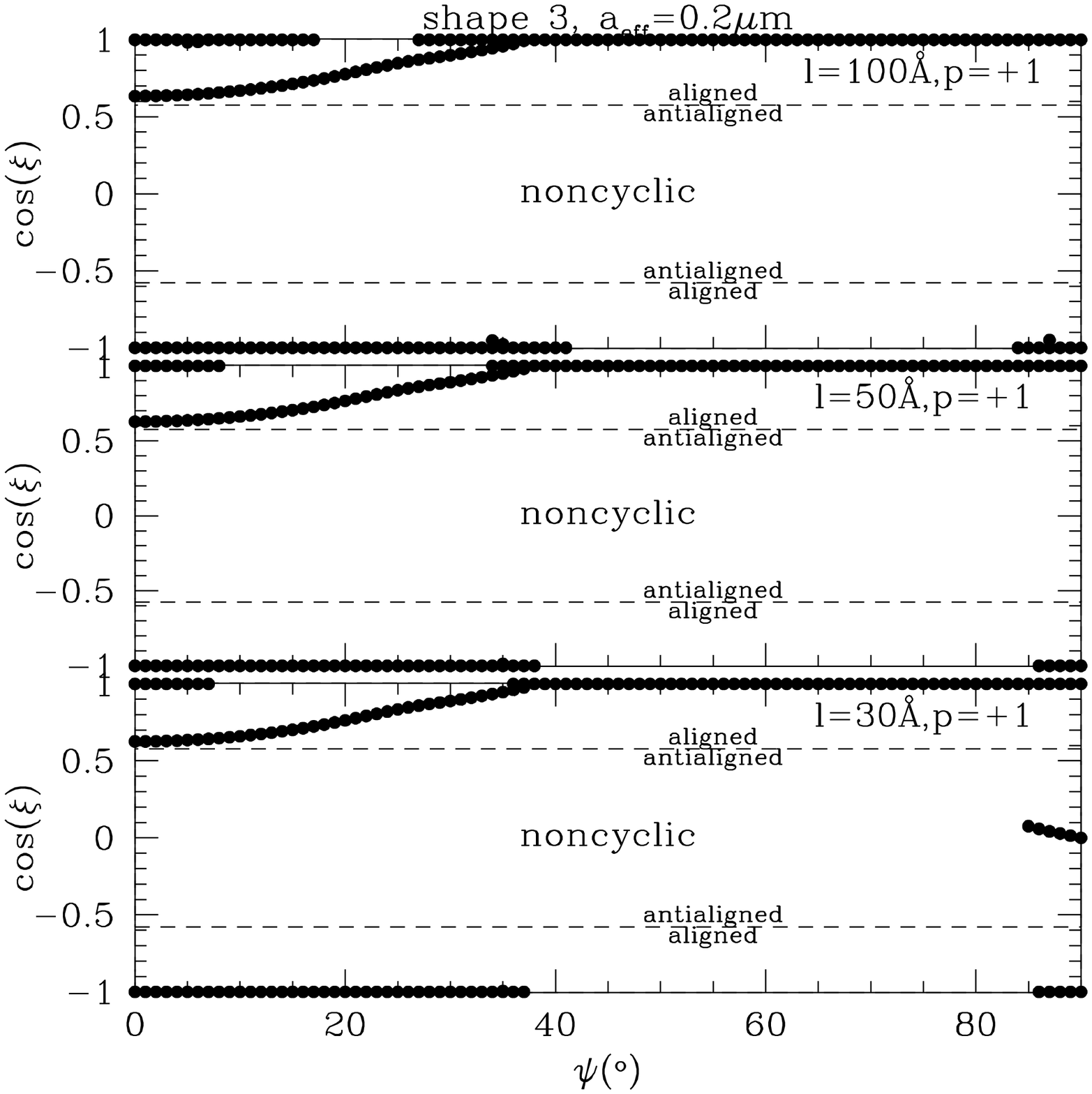}
\caption{
	\label{fig:omega.g3p3510}
	Same as fig. \protect\ref{fig:omega.g3} but for $\HH$ formation
	torques with $p=1$ and $l= 30, 50, 100\Angstrom$.
	}
\end{figure}

\begin{figure}
\epsscale{0.5}
\plotone{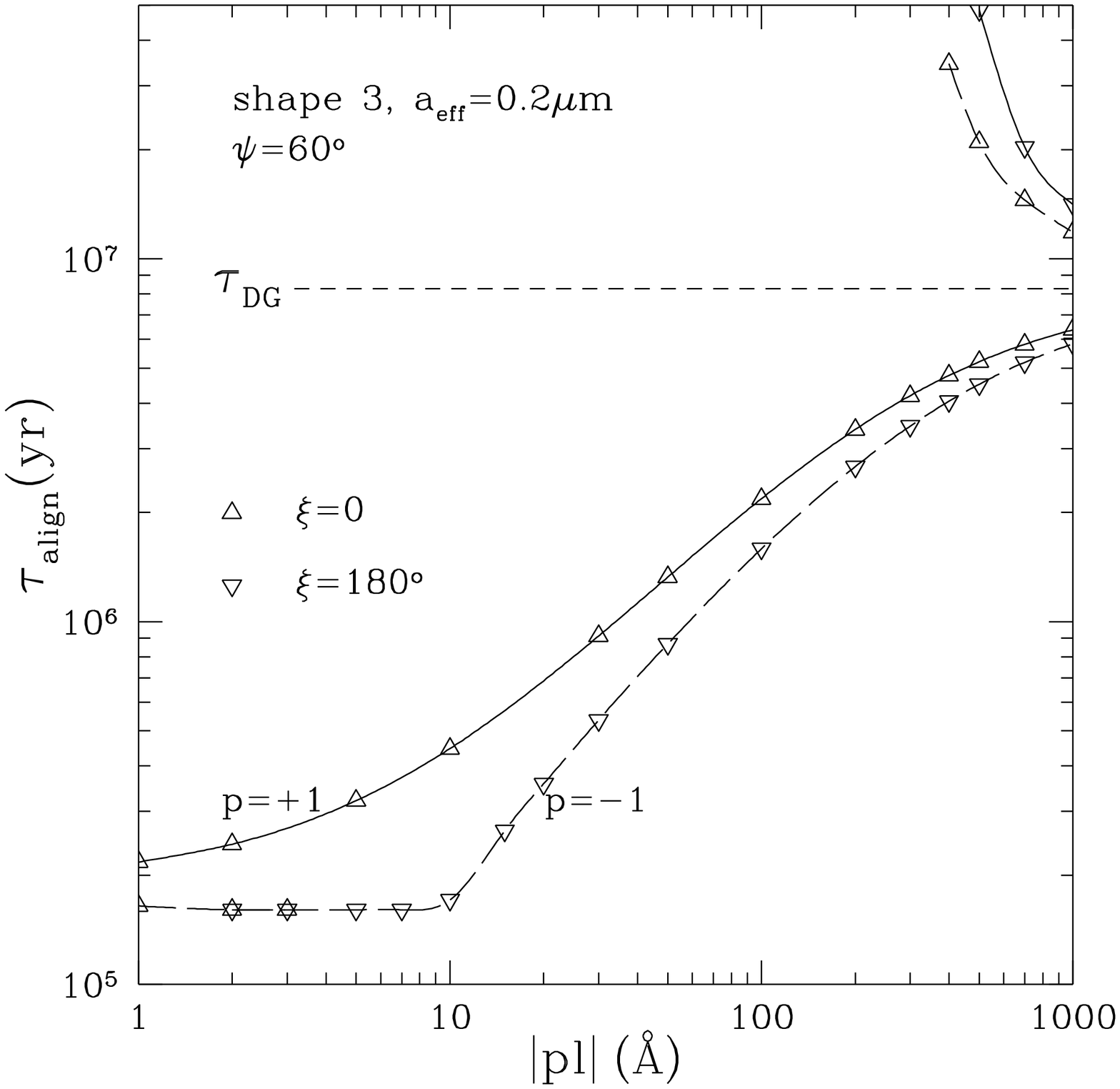}
\caption{
	\label{fig:talign}
	The alignment time for $a_\eff=0.2\micron$ grain with shape 3 of 
	Fig.\ \protect{\ref{fig:shapes}}, for angle $\psi=60\arcdeg$ 
	between $\bB_0$ and the starlight anisotropy, as a function
	of the surface length scale $l$, where the surface density of
	active $\HH$ formation sites is $l^{-2}$.
	The alignment time is shown for $\HH$ 
	formation torque given by eq.\ \protect{\ref{eq:gamma_h2}},
	directed parallel ($p=+1$) or antiparallel ($p=-1$) to the
	grain axis $\ahat_1$.
	We see that for $l<10^3\Angstrom$, the alignment time is
	significantly smaller than the timescale $\tau_\DG$ for
	alignment by paramagnetic dissipation.
	For $l<50\Angstrom$, grain alignment takes place in $\ltsim10^6\yr$.
	Note that for $l\gtsim400\Angstrom$ two alignment times are shown,
	corresponding to $\xi=0$ and $180\arcdeg$; for $\xi=0$ the radiative
	torques assist alignment, while for $\xi=180\arcdeg$ paramagnetic
	alignment must overcome the radiative torques.
	}
\end{figure}

The numerical examples considered above assume 
the characteristic separation between $\HH$ formation sites to
be $l=10\Angstrom$,
with the $\HH$ formation torque
given by eq.\ \ref{eq:gamma_h2} with $p=0,\pm1$.
It is quite possible that the surface density of active sites is lower
($l>10\Angstrom$), in which case the r.m.s. $\HH$ torque will be larger.
To explore this, we consider shape 3 with $a=0.2\micron$.
In Figure \ref{fig:omega.g3p3510} we show the stationary solutions for
$pl=\pm30,50,100\Angstrom$.
For $|pl|\gtsim20\Angstrom$ the $\HH$ formation torque dominates the
grain spinup, and we
no longer have cyclic trajectories (all crossovers have the same polarity).
However, we still have multiple stationary points in many cases.
The values of $R$ for these cases do not appear to be very sensitive to the
value of $pl$.
Furthermore, even though the $\HH$ formation torques may dominate radiative
torques in driving superthermal rotation, radiative torques may still be
more important than Davis-Greenstein torques in bringing about alignment
of the grain angular momentum with $\bB_0$.
This is evident in Figure \ref{fig:talign}, showing the alignment time
as a function of $l$ for $\psi=60\arcdeg$.
Since the Davis-Greenstein torque is proportional to $\omega$, in the limit
of large $l$ the alignment time approaches the Davis-Greenstein alignment
time.
However, if the active sites for $\HH$ formation are characterized by
$l\ltsim300\Angstrom$ (i.e., $>1$ active site per $10^5\Angstrom^2$) then
radiative torques apparently dominate the grain alignment process.

\subsection{Cyclic Maps}

\begin{figure}
\epsscale{0.5}
\plotone{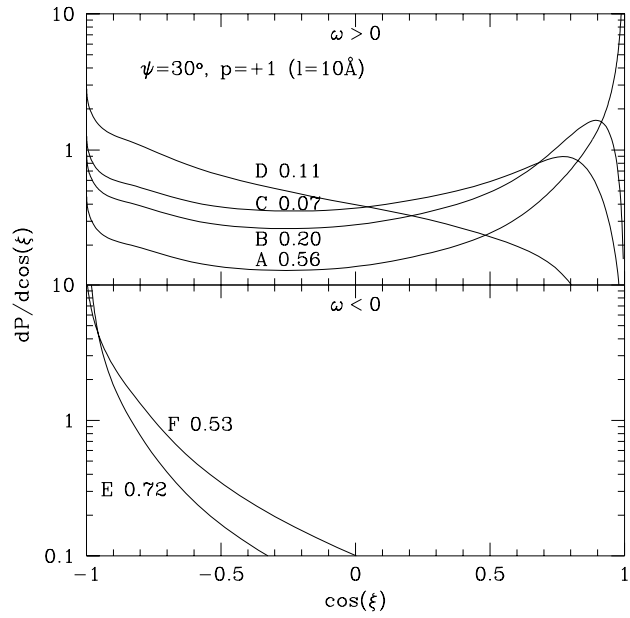}
\caption{
	\label{fig:dpdcosxi}
	Plot of the fractional time distribution over $\cos\xi$ for
	selected trajectories beginning near one crossover point and ending at
	the other, for an $a_\eff=0.2\micron$ grain with shape 1 and
	$\psi=30\arcdeg$ (see Fig. \protect{\ref{fig:tr_30.g1p}}).
	Distributions are shown for 6 trajectories indicated by letters
	A--F in 
	Fig.\protect{\ref{fig:tr_30.g1p}}.
	For each trajectory the value of $\langle R\rangle$ is given.
	}
\end{figure}

When the trajectory map shows cyclic behavior, the degree of grain alignment
will depend on what fraction of the time the grain spends with different
values of $\xi$.
Suppose we have two crossover attractors, at $\xi_{c1}$ and $\xi_{c2}$,
with polarities $+$ and $-$, respectively.
Consider a trajectory beginning near $\xi_{c1}$, at
$\xi_{c1}+\delta\xi$ and
$\omega=\delta\omega$.
The total time to reach the next crossover is
\beq
\Delta t = \int_{\xi_{c1}+\delta\xi}^{\xi_{c2}} d\xi \left( d\xi/dt\right)^{-1}
~~~.
\eeq
The fraction $dP$ of the time spent in an interval
$d\cos\xi$ is given by
\beq
{dP\over d\cos\xi} = {1\over \Delta t\sin\xi|d\xi/dt|}
~~~.
\eeq
In Figure \ref{fig:dpdcosxi} we show the distribution computed for 
6 trajectories for grain shape 1 and
$\psi=30\arcdeg$ (see Fig. \ref{fig:tr_30.g1p}).
The upper panel shows four trajectories emerging from the crossover
attractor at $\cos\xi=1$; the lower panel is for two trajectories emerging
from the crossover attractor at $\cos\xi=-1$.
For each trajectory we compute the time-averaged 
alignment factor
\beq
\langle R \rangle = \int_{-1}^1 d\cos\xi {dP\over d\cos\xi} R(\xi)~~~.
\eeq
We see that quite different values of $\langle R\rangle$ are found for the
different trajectories, but in this case at least there is a tendency for
the average alignment
$\langle R\rangle$ to be positive.
Thus, although we are not yet able to predict which trajectory (or
distribution of trajectories) will be taken by grains emerging from
the crossover attractors, there is at least an indication that
even cyclic behavior of the kind shown in Fig.\ \ref{fig:tr_30.g1p}
may result in significant grain alignment.

\subsection{Future Work}

The present paper is an exploratory study of the role of starlight torques
in the dynamics and alignment of interstellar grains.
Three grain geometries have been studied, but for only one size,
$a_\eff=0.2\micron$.
The important role of starlight torques has been demonstrated, and we
conclude that $a\approx0.2\micron$ grains can often be effectively aligned with
the magnetic field $\bB_0$ by this process.

A number of issues remain to be investigated:
\begin{enumerate}
\item It is important to examine the dependence of this mechanism on the
grain size.  In Paper I we showed that radiative torques are 
relatively unimportant
for $a_\eff\ltsim0.05\micron$ grains with the geometry of shape 1;
based on those results it seems likely that radiative torques will not
be able to produce alignment of $a\ltsim0.1\micron$ grains, but
future work should examine the dependence on grain size
in greater detail, with the important goal of understanding the
observed minimal alignment of grains with $a\ltsim0.1\micron$
(Kim \& Martin 1995).
\item The present study has been restricted to the dielectric function of
silicate material.  It will be of interest to determine whether other 
grain materials, e.g., graphite, will behave similarly.
\item We have not attempted to examine the grain dynamics when the
rate of rotation approaches ``thermal'' values during the ``crossover''
process.  A detailed study of the crossover process is essential for
understanding the fate of grains having cyclic or semicyclic trajectory
maps.
\end{enumerate}
These issues will be addressed in future investigations.

\section{Summary\label{sec:summary}}

Our principal results are as follows:
\begin{enumerate}
\item As anticipated in Paper I, anisotropic starlight incident on an
interstellar grain
produces a torque which, in addition to spinning the grain up to 
superthermal rotation rates, directly 
affects the orientation of the grain
relative to both the anisotropy direction and the direction of the
local magnetic field $\bB_0$.
Accordingly, previous studies of grain alignment, which neglected these
torques, are incomplete.

\item We obtain equations of motion for the grain which are valid 
in the limit where the grain rotation period and the 
precession period are both short compared to the timescale for
the grain to change its angular velocity and orientation relative
to the magnetic field.

\item The equations of motion have been integrated for three different
grains and many different initial conditions.
For a given grain and a given angle $\psi$ between the 
radiation anisotropy direction and
$\bB_0$, we produce a ``trajectory map'' which
shows how the grain will evolve under the combined effects of starlight,
$\HH$ formation torques, gas drag, and paramagnetic dissipation
(see Figs. \ref{fig:tr_00.g1z}--\ref{fig:tr_00.g3p}).

\item Under some conditions the trajectory maps contain a ``stationary
attractor''; trajectories coming near this point are permanenty captured.
The stationary attractor often (but not always) corresponds to a state
of perfect alignment of the grain angular momentum $\bJ$ with the magnetic 
field $\bB_0$.

\item The trajectory map may contain one or more ``crossover attractors''
where (formally) the grain rotation may reverse.
Unfortunately, the assumptions underlying our dynamical study 
break down near such crossover points, and the present study is
unable to describe the dynamical evolution of the grain near the crossover
point.

\item Trajectory maps can be classified as ``noncyclic'', ``cyclic'', and
``semicyclic'', based upon whether or not there are stationary attractors,
and upon the locations and polarities of crossover attractors, and the
locations of crossover repellors.

\item In diffuse clouds, for grains with effective radii 
$a_\eff\gtsim0.1\micron$,
these starlight torques dominate the grain dynamics, if the
separation between active sites for $\HH$ formation is $l\ltsim20\Angstrom$.
Even when $\HH$ torques dominate the grain spinup ($l\gtsim20\Angstrom$)
the radiation torques dominate the grain alignment
process for $l\ltsim300\Angstrom$.

\item Based on study of three grain geometries, we conclude that
radiative torques due to anisotropic starlight appear to be the primary 
mechanism
responsible for the observed alignment of interstellar dust grains.
\end{enumerate}
\acknowledgements
We are grateful to R.H. Lupton for the availability of the SM plotting
package,
and to
A. Lazarian and L. Spitzer, Jr., for
helpful discussions.
This research was supported in part by NSF grant
AST-9219283 to BTD, and by an NSF Fellowship
to JCW.
\appendix
\section{Appendix: The Functions $F(\xi,\phi)$, $G(\xi,\phi)$, and
$H(\xi,\phi)$}

The expressions (\ref{eq:F(xi,phi)}--\ref{eq:H(xi,phi)}) for
$F(\xi,\phi)$, $G(\xi,\phi)$, and
$H(\xi,\phi)$ are given in terms of the torque efficiency vectors
$\langle\bQ_\Gamma(\Theta,\Phi)\rangle$.
As discussed in Paper I (eq. 41), 
it is possible to obtain $\bQ_\Gamma(\Theta,\Phi)$
from $\bQ_\Gamma(\Theta,0)$, so that the time-consuming 
calculations required to obtain $\bQ_\Gamma$ can be restricted to $\Phi=0$.
Thus, we obtain $\Theta(\xi,\phi)$ and $\Phi(\xi,\phi)$
from eq.(\ref{eq:Theta},\ref{eq:Phi}), and then obtain
$F(\xi,\phi)$, $G(\xi,\phi)$ and $H(\xi,\phi)$ from
$\bQ_\Gamma(\Theta,0)$:
\begin{eqnarray}
F(\xi,\phi)
&=~&
\langle\bQ_\Gamma(\Theta,0)\rangle\!\cdot\!\ehat_1
\left[-\sin\!\psi\cos\!\xi\cos\!\phi-\cos\!\psi\sin\!\xi\right]
\nonumber
\\
&~+&
\langle\bQ_\Gamma(\Theta,0)\rangle\!\cdot\!\ehat_2
\left[
\cos\Phi(\cos\psi\cos\xi\cos\phi-\sin\psi\sin\xi) + \sin\Phi\cos\xi\sin\phi
\right]
\nonumber
\\
&~+&
\langle\bQ_\Gamma(\Theta,0)\rangle\!\cdot\!\ehat_3
\left[
\cos\Phi\cos\xi\sin\phi + \sin\Phi(\sin\psi\sin\xi-\cos\psi\cos\xi\cos\phi)
\right]
\\
G(\xi,\phi)
&=~&
\langle\bQ_\Gamma(\Theta,0)\rangle\!\cdot\!\ehat_1
\left[\sin\!\psi\sin\!\phi\right]
\nonumber
\\
&~+&\langle\bQ_\Gamma(\Theta,0)\rangle\!\cdot\!\ehat_2
\left[\sin\Phi\cos\phi-\cos\Phi\cos\psi\sin\phi\right]
\nonumber
\\
&~+&
\langle\bQ_\Gamma(\Theta,0)\rangle\!\cdot\!\ehat_3
\left[\cos\Phi\cos\phi+\sin\Phi\cos\psi\sin\phi\right]
\\
H(\xi,\phi)
&=~&
\langle\bQ_\Gamma(\Theta,0)\rangle\!\cdot\!\ehat_1
\left[\cos\!\psi\cos\!\xi- \sin\!\psi\sin\!\xi\cos\!\phi\right]
\nonumber
\\
&~+&
\langle\bQ_\Gamma(\Theta,0)\rangle\!\cdot\!\ehat_2
\left[
\cos\Phi(\sin\psi\cos\xi+\cos\psi\sin\xi\cos\phi)+\sin\Phi\sin\xi\sin\phi
\right] ~~~.
\end{eqnarray}

%************************
\begin{deluxetable}{cccc}
\tablecaption{Coordinates of Constituent Blocks for Shape 2\label{tab:shape2}}
\tablehead{
\colhead{$j$}&
\colhead{$x_j$}&
\colhead{$y_j$}&
\colhead{$z_j$}
	}
\startdata
1&0&0&0\nl
2&1&0&0\nl
3&0&1&0\nl
4&1&1&0\nl
5&0&0&1\nl
6&1&0&1\nl
7&0&1&1\nl
8&1&1&1\nl
9&2&0&0\nl
10&2&1&0\nl
11&0&0&2\nl
\enddata
\end{deluxetable}

\begin{deluxetable}{ccccc}
\tablecaption{Coordinates of Constituent Spheres for Shape 3\label{tab:shape3}}
\tablehead{
\colhead{$j$}&
\colhead{$x_j$}&
\colhead{$y_j$}&
\colhead{$z_j$}&
\colhead{$r_j$}
	}
\startdata
1&1.88	&-0.87	&1.54	&4.22	\nl
2&-2.20	&1.89	&0.14	&3.86	\nl
3&-2.23	&0.33	&-2.15	&3.46	\nl
4&0.57	&1.20	&0.38	&4.98	\nl
5&5.56	&-1.06	&-0.68	&4.06	\nl
\enddata
\end{deluxetable}

\begin{deluxetable}{cccccc}
\tablecaption{Principal Axes $\ahat_j$ and factors $\alpha_j$ for
Shapes 2 and 3\label{tab:alphaj}}
\tablehead{
\colhead{shape}&
\colhead{$j$}&
\colhead{$(\ahat_j)_x$}&
\colhead{$(\ahat_j)_y$}&
\colhead{$(\ahat_j)_z$}&
\colhead{$\alpha_j$}
	}
\startdata
2&1&0.2273&0.8398&0.4930	&1.561\nl
2&2&0.5681&-0.5256&0.6333	&1.464\nl
2&3&0.7909&0.1361&-0.5965	&0.889\nl
3&1&0.3435&0.9357&-0.0802	&1.378\nl
3&2&0.0222&0.0773&0.9967	&1.332\nl
3&3&0.9389&-0.3441&0.0058	&0.765\nl
\enddata
\end{deluxetable}

\begin{deluxetable}{ccc}
\tablecaption{
	Radiative Torque Efficiency Factors for Shapes 1--3, 
	$a_\eff=0.2\micron$
	\label{tab:qfactors}
	}
\tablehead{
	\colhead{shape}&
	\colhead{$\langle Q_\Gamma^{iso}\rangle_{ISRF}$}&
	\colhead{$\ahat_1\cdot\langle\bQ_{\Gamma}\rangle_{ISRF} (\Theta=0)$}
	}
\startdata
1	&$ 1.99\times10^{-4}$	&$ 7.01\times10^{-2}$\nl
2	&$-1.10\times10^{-4}$	&$-2.46\times10^{-2}$\nl
3	&$ 2.38\times10^{-4}$	&$-1.39\times10^{-2}$\nl
\enddata
\end{deluxetable}

\clearpage
\clearpage
\end{document}